\documentclass{article}

\usepackage{PRIMEarxiv}

\usepackage[utf8]{inputenc} 
\usepackage[T1]{fontenc}    
\usepackage{hyperref}       
\usepackage{url}            
\usepackage{booktabs}       
\usepackage{amsfonts}       
\usepackage{nicefrac}       
\usepackage{microtype}      
\usepackage{lipsum}
\usepackage{fancyhdr}       
\usepackage{graphicx}       
\graphicspath{{media/}}     
\usepackage{siunitx}
\usepackage{booktabs}
\usepackage[caption=false,labelformat=simple]{subfig}
\usepackage{flafter} 
\usepackage{placeins}
\usepackage{authblk}
\usepackage{amsmath,amssymb}
\usepackage[font=small,labelfont=bf]{caption}
\ifdefined\qty\else
  \ifdefined\NewCommandCopy
    \NewCommandCopy\qty\SI
  \else
    \NewDocumentCommand\qty{O{}mm}{\SI[#1]{#2}{#3}}
  \fi
\fi
\ifdefined\unit\else
  \ifdefined\NewCommandCopy
    \NewCommandCopy\unit\si
  \else
    \NewDocumentCommand\unit{O{}m}{\si[#1]{#2}}
  \fi
\fi

\makeatletter
\LetLtxMacro\@uthorfrom@uthblk\author

\RenewDocumentCommand{\author}{+o+m}{%
  \ifnum0=\value{authors}%
  \def\AB@authors{}%
  \fi
  \@uthorfrom@uthblk[#1]{#2}%
}
\makeatother

\makeatletter
\renewcommand\AB@affilsepx{, \protect\Affilfont}
\makeatother

\title{\boldmath Software Compensation for Highly Granular Calorimeters using Machine Learning}

\author[ ]{\begin{Large}\begin{center}The CALICE Collaboration\end{center}\end{Large}}


\author[a]{S.\,Lai,}
\author[a]{J.\,Utehs,}
\author[a]{A.\,Wilhahn}

\author[b]{O.\,Bach,}
\author[b]{E.\,Brianne,}
\author[b]{A.\,Ebrahimi,} 
\author[b]{K.\,Gadow,}
\author[b]{P.\,G\"{o}ttlicher,}
\author[b,1]{O.\,Hartbrich,\thanks{{now at Oak Ridge National Laboratory, 1 Bethel Valley Road, Oak Ridge, TN 37830, USA}}}
\author[b]{D.\,Heuchel,}
\author[b,2]{A.\,Irles, \thanks{now at Instituto de Física Corpuscular, Parque Científico, Catedrático José Beltrán, 2 | E-46980 Paterna, España}}
\author[b]{K.\,Kr\"{u}ger,}
\author[b,3]{J.\,Kvasnicka \thanks{also at Institute of Physics, The Czech Academy of Sciences}}
\author[b]{ S.\,Lu,}
\author[b]{C.\,Neub\"{u}ser,}
\author[b]{A.\,Provenza,}
\author[b]{M.\,Reinecke,}
\author[b]{F.\,Sefkow,}
\author[b,4]{S.\,Schuwalow,\thanks{deceased}}
\author[b]{M.\,De Silva,}
\author[b]{Y.\,Sudo,}
\author[b]{H.L.\,Tran,}

\author[c]{E.\, Buhmann,}
\author[c]{E.\,Garutti,}
\author[c]{S.\.Huck,}
\author[c]{G.\,Kasieczka,}
\author[c]{S.\,Martens,}
\author[c,5]{J.\,Rolph\thanks{corresponding author}}
\author[c]{J.\, Wellhausen}

\author[d]{G. C.\,Blazey} 
\author[d]{A.\,Dyshkant,}
\author[d]{K.\,Francis,}
\author[d]{V.\,Zutshi,}

\author[e]{B.\,Bilki,}
\author[e]{D.\,Northacker,}
\author[e]{Y.\,Onel,}

\author[f]{F.\,Hummer,}
\author[f]{F.\,Simon,}

\author[g]{K.\,Kawagoe,}
\author[g]{T.\,Onoe,}
\author[g, 6]{T.\,Suehara\thanks{now at ICEPP, The University of Tokyo, 7-3-1 Hongo, Bunkyo-ku, Tokyo},}
\author[g]{S.\,Tsumura,}
\author[g]{T.\,Yoshioka,}

\author[h]{M.C.\,Fouz,}

\author[i]{L.\,Emberger,}
\author[i]{C.\,Graf,}
\author[i]{M.\,Wagner,}

\author[j]{R.\,P\"oschl,}
\author[j]{F.\,Richard,}
\author[j]{D.\,Zerwas,}

\author[k]{V.\,Boudry,}
\author[k]{J-C.\,Brient,}
\author[k]{J.\,Nanni,}
\author[k]{H.\,Videau,}

\author[l]{L.\,Liu,}
\author[l]{R.\,Masuda,}
\author[l]{T.\,Murata,}
\author[l]{W.\,Ootani,}
\author[l]{T.\,Takatsu,}
\author[l]{N.\,Tsuji,}

\author[m]{M.\,Chadeeva,}
\author[m]{M.\,Danilov,}
\author[m]{S.\,Korpachev,}
\author[m]{V.\,Rusinov}

\affil[a]{II. Physikalisches Institut, Georg-August-Universit\"at G\"ottingen, Friedrich-Hund-Platz 1, D-37077 G\"ottingen, Germany}

\affil[b]{DESY, Notkestrasse 85, D-22603 Hamburg, Germany}

\affil[c]{Univ. Hamburg, Physics Department, Institut f\"ur Experimentalphysik, Luruper Chaussee 149, 22761 Hamburg, Germany}

\affil[d]{NICADD, Northern Illinois University, Department of Physics, DeKalb, IL 60115, USA}

\affil[e]{University of Iowa, Dept. of Physics and Astronomy, 203 Van Allen Hall, Iowa City, IA 52242-1479, USA}

\affil[f]{Karlsruhe Institute of Technology, Institute for Data Processing and Electronics, Kaiserstr. 12, D-76131 Karlsruhe, Germany}

\affil[g]{Department of Physics and Research Center for Advanced Particle Physics, Kyushu University, 744 Motooka, Nishi-ku, Fukuoka 819-0395, Japan}

\affil[h]{CIEMAT, Centro de Investigaciones Energeticas, Medioambientales y Tecnologicas, Madrid, Spain}

\affil[i]{Max-Planck-Institut f\"ur Physik, F\"ohringer Ring 6, D-80805 Munich, Germany}

\affil[j]{Université Paris-Saclay, CNRS/IN2P3, IJCLab, 91405 Orsay, France}

\affil[k]{Laboratoire Leprince-Ringuet (LLR), CNRS, \'{E}cole polytechnique, Institut Polytechnique de Paris, F-91120 Palaiseau, France}

\affil[l]{ICEPP, The University of Tokyo, 7-3-1 Hongo, Bunkyo-ku, Tokyo 113-0033, Japan}

\affil[m]{Affiliated with an institute that has signed the CALICE MOU}

\begin{document}

\maketitle

\begin{abstract}
A neural network for software compensation was developed for the highly granular CALICE Analogue Hadronic Calorimeter (AHCAL). The neural network uses spatial and temporal event information from the AHCAL and energy information, which is expected to improve sensitivity to shower development and the neutron fraction of the hadron shower. The neural network method produced a depth-dependent energy weighting and a time-dependent threshold for enhancing energy deposits consistent with the timescale of evaporation neutrons. Additionally, it was observed to learn an energy-weighting indicative of longitudinal leakage correction. In addition, the method produced a linear detector response and outperformed a published control method regarding resolution for every particle energy studied.
\end{abstract}

\section{Introduction}

To fulfil the requirements for BSM physics searches and Higgs precision measurements at future linear colliders, a challenging final state jet-energy resolution must be achieved. For example, for ILC operating at $\sqrt{s}$= 0.5$-$\qty{1}{\tera \electronvolt} where typical di-jet energies for interesting physics processes will be in the range 150–\qty{350}{\giga \electronvolt}, a jet energy resolution of \qty{2.7}{\percent} is crucial \cite{particle_flow}. Particle Flow (PF) is a method expected to provide this resolution, which relies upon accurate tracking of charged particles in a jet, sophisticated event reconstruction techniques, and highly granular sampling calorimeters. A prototype of such a detector is the CALICE Analogue Hadronic Calorimeter (AHCAL) \cite{AHCAL}, a highly-granular steel-scintillator sampling calorimeter designed for PF, with $24\times24\times38$ individual silicon photomultiplier (SiPM) readout cells. The AHCAL is notable for its capacity to measure a timestamp for each readout channel.

 The response of calorimeters to hadrons may be described in terms of two components: an electromagnetic component (produced mainly by $\pi^{0}/\eta \rightarrow \gamma\gamma$, contributed to by nuclear $\gamma$), and a hadronic component, which contains the remainder of energy depositing processes. The calorimeter response is therefore split into an EM response ($e$) and a HAD response ($h$).  A hadron shower in a calorimeter exhibits an EM-dominated, energy-dense ’core’ that propagates over a short longitudinal and lateral range and a HAD-dominated, diffuse energy-sparse ’halo’, which propagates over a wider range  \cite{wigmans}. Part of the energy deposited by a hadron shower cannot be detected and is called ’invisible energy’ (e.g. neutrinos, nuclear binding energy losses). This fraction also experiences significant stochastic fluctuations from event to event, contributing to the calorimeter's resolution.

 Compensation describes a method to equalise $e$ and $h$, typically by attenuating $e$ and enhancing $h$ to improve the resolution. Hardware compensation requires careful tuning of the composition and proportions of active and passive material in the calorimeter. This method is difficult to implement in highly-granular calorimeters, which require a high degree of longitudinal segmentation. Therefore, software compensation (SC) algorithms are employed for this purpose and operate by estimating the EM fraction of a shower using information measured in each event. 
 
 Notably, spatial and temporal readout information available from highly granular calorimeter may be used for SC: 
 
 \begin{itemize}
     \item  A highly granular calorimeter may be able to resolve the hadron shower core and halo, and therefore exploit spatial energy density for SC;
     
     \item The number of neutrons produced in nuclear interactions is proportional, on average, to the invisible energy of the hadron shower. Energy deposits from neutrons can be measured indirectly using ionisation by recoil protons from neutron elastic scattering in hydrogenous active material such as plastic scintillator and photons from neutron capture. Energy deposits induced by neutrons are delayed by $10$-$\qty{100}{\nano \second}$ in steel \cite{time}. A time-sensitive hadron calorimeter may therefore exploit temporal information for SC. 
 \end{itemize}
  
Artificial neural network models have already been demonstrated to effectively exploit the spatial development of hadron showers to improve SC. For example, a study performed in Ref.~\cite{FCChh} demonstrated that a deep neural network was found to improve the response of a highly-granular hadron calorimeter system from  $\qty{48}{\percent}/\sqrt{E_{\mathrm{particle}}} \oplus \qty{2.2}{\percent}$ to $\qty{37}{\percent}/\sqrt{E_{\mathrm{particle}}} \oplus \qty{1}{\percent}$ using simulation.  

However, a similar studies performed for AHCAL in Ref.~\cite{erik} and Ref.~\cite{clusterSC}, which trained and compared the performance of neural networks trained on both simulation and testbeam data, demonstrated the inability of similar machine learning-based SC algorithms to interpolate or extrapolate compensation from the limited hadron shower data typically available for such studies. In other words, the SC algorithm was biased to the training range of energies and its binning. This result is problematic as it indicates that experimental data of hadron showers from testbeams cannot be used for training SC algorithms because the available samples are typically binned too coarsely in particle energy to prevent bias. Additionally, while simulation samples can be used to train the algorithm with no constraints on particle energy binning or ranges, producing and storing these samples is presently an unsustainable practice. These limitations therefore motivate the development of an algorithm that can exploit the spatial and temporal information from AHCAL and simultaneously remain unbiased to the training particle energies. 
  
In the presented study, a neural network is designed to perform SC on simulated $\pi^{-}$ hadron showers observed with the AHCAL calorimeter, using the local spatial and temporal energy density from the event rather than just the sum of energy deposits. This information was expected to reduce the effect of stochastic fluctuations by improving sensitivity to the shower development and the neutron fraction of the event. Importantly, the neural network was carefully structured to reduce the effect of energy biasing. Finally, the neural network is compared to the standard CALICE SC method, which is used as a control algorithm. The results are then compared.

\section{Methods and Tools}

\label{sec:CALICE_AHCAL}

\begin{figure}[!h]
\centering
\subfloat[%
 \label{fig:Physics_AHCAL}
]{\includegraphics[width=0.49\linewidth]{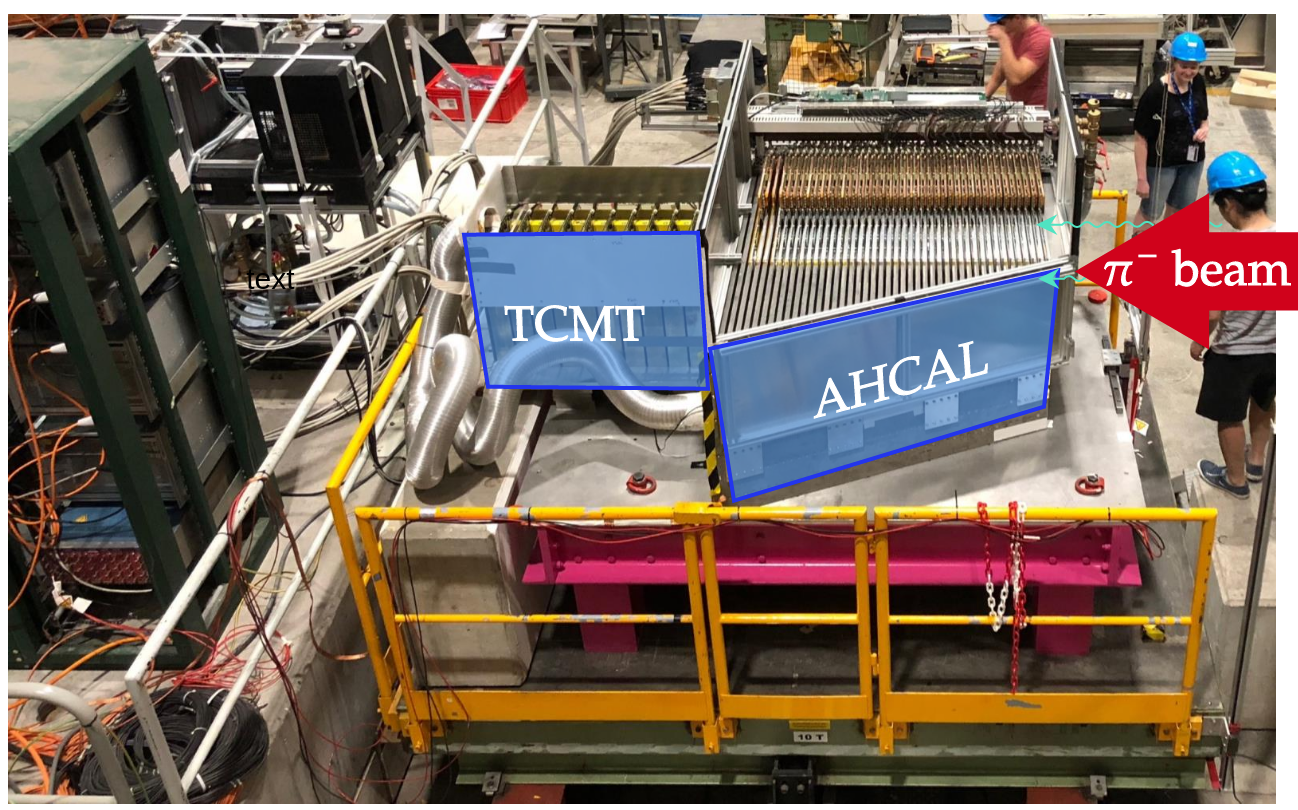}}
\hfill
\subfloat[%
 \label{fig:Physics_AHCAL_IJ}
]{\includegraphics[width=0.44\linewidth]{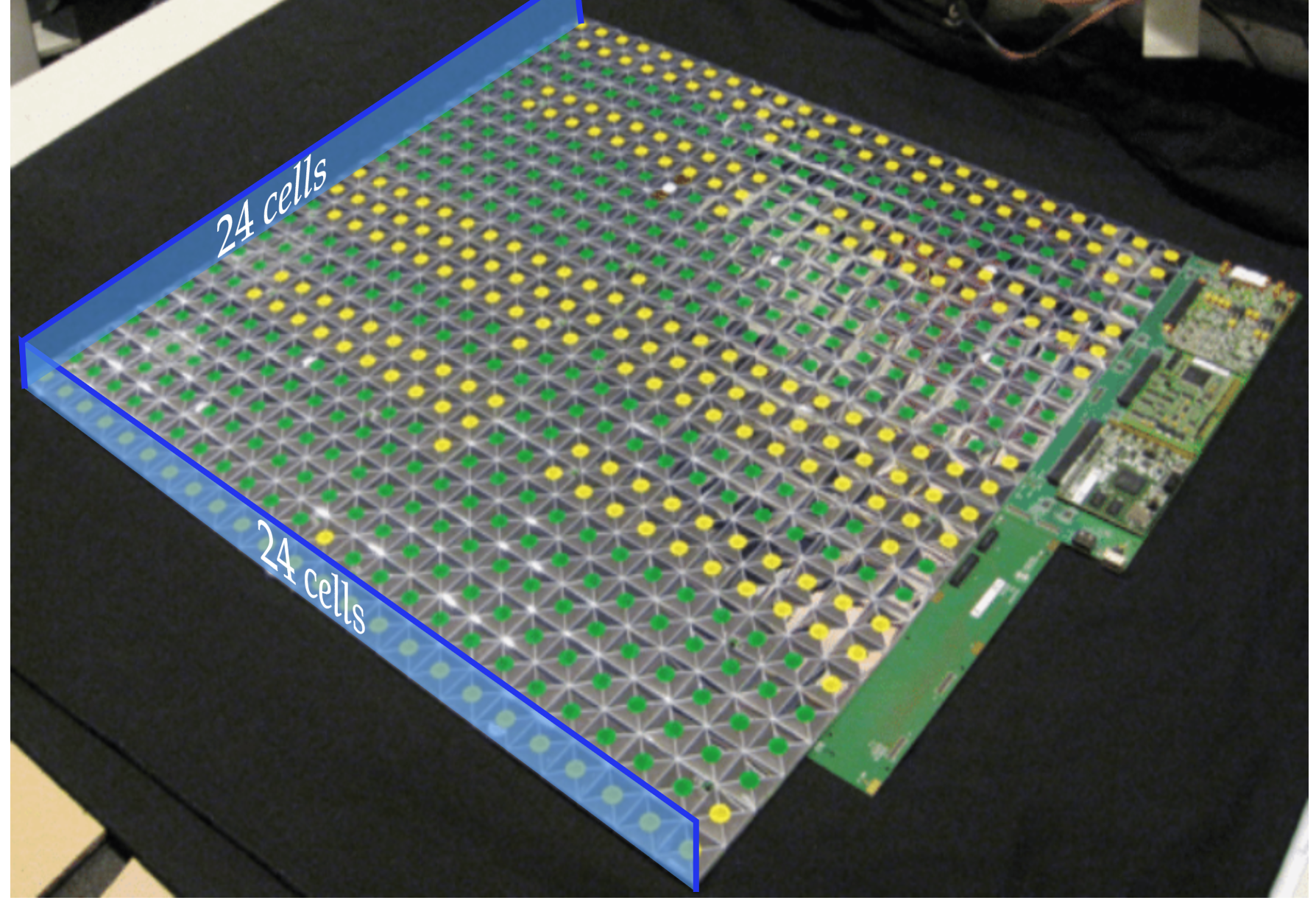}}

\caption{Pictures showing the CALICE AHCAL at testbeam. Fig.~\ref{fig:Physics_AHCAL} shows the detector setup for a testbeam performed in June 2018 at the Super Proton Synchrotron (SPS) at CERN, Geneva \cite{Testbeam}. Fig.~\ref{fig:Physics_AHCAL_IJ} shows the individual cells of the calorimeter wrapped in foil to improve photon sensitivity.}
\label{fig:AHCAL}
\end{figure}

The CALICE AHCAL is a non-compensating, highly granular steel-scintillator calorimeter prototype designed for future precision $\mathrm{e}^{+}$-$\mathrm{e}^{-}$ collider experiments. It has a highly granular structure, consisting of $24\times24\times38$ plastic scintillator cells of $30\times30\times \qty{3}{\milli \meter \cubed}$ volume each, read out by SiPMs. These cells indicate the spatial position, magnitude and timestamp of energy deposition with a minimum operating time resolution of up to \qty{100}{\pico \second} allowed by hardware. The detector has a depth of approximately $4.2$ nuclear interaction lengths ($\mathrm{\lambda}_{\mathrm{I}}$). The hadronic calorimeter is complemented by a steel-scintillator Tail Catcher/Muon Tracker (TCMT) detector, composed of $320$ extruded scintillator strips of $50 \times \qty{5}{\milli \meter \squared}$ area packaged in $16 \times \qty{1}{\meter \squared}$ planes interleaved between steel plates corresponding to an additional depth of \qty{1.1}{\text{\ensuremath{\lambda_{\mathrm{I}}}}} \cite{TCMT}. The TCMT is used in this analysis to tag leakage. Pictures of the AHCAL calorimeter are shown for reference in Fig.~\ref{fig:AHCAL}.

Event information from AHCAL consists of the position of an active cell with an energy deposit in the AHCAL cell matrix ($I_{\mathrm{hit}}$, $J_{\mathrm{hit}}$, $K_{\mathrm{hit}}$), its energy in calibrated MIP units ($E_{\mathrm{hit}}$), and its timestamp in nanoseconds, relative to the time at which deposited energy in a given cell cross a pre-defined threshold ($t_{\mathrm{hit}}$). $I_{\mathrm{hit}}$ and $J_{\mathrm{hit}}$ indicate the lateral spatial position of an active cell relative to the longitudinal axis of the calorimeter ($I_{\mathrm{hit}}$, $J_{\mathrm{hit}} \in \left[1, 24\right]$ in units of cell index). The longitudinal spatial position (depth in layers) is denoted $K_{\mathrm{hit}}$ ($K_{\mathrm{hit}} \in \left[1, 38\right]$ in units of layer index). The energy of an active cell is denoted $E_{\mathrm{hit}}$, measured in Analogue-to-Digital counts, calibrated to the energy deposited by a minimum ionising particle (MIP) in one cell \cite{AHCALCalib}. $E_{\mathrm{hit}}$ takes a value between a noise threshold at \qty{0.5}{\text{MIP}} and the energy corresponding to the SiPM saturation value. The $t_{\mathrm{hit}}$ is bounded between the time at which the energy deposited in a given cell crosses a pre-defined threshold (normalised to \qty{0}{\nano \second} in this study), smeared by the resolution, and the chosen gate length for the measurement of an event. This study considers the ultimate \qty{100}{\pico \second} timing resolution for AHCAL. No charge integration gate length is considered in this study. The calorimeter response is measured as the sum of the individual active cells (hits) in an event, $E_{\mathrm{sum}} = \sum^{\mathrm{event}} E_{\mathrm{hit}}$. Additionally, the incident position of a charged particle in lateral coordinates is reconstructed using four delay wire chambers (DWC) of $10 \times \qty{10}{\centi \meter \squared}$ size, which is denoted as a vector $\left[ I_{\mathrm{track}}, J_{\mathrm{track}} \right]$ \cite{Testbeam}. The track information is only relevant to event selection cuts described in Section \ref{sec:Datasets}.

\begin{figure}[!h]
\centering
\includegraphics[width=0.6\linewidth]{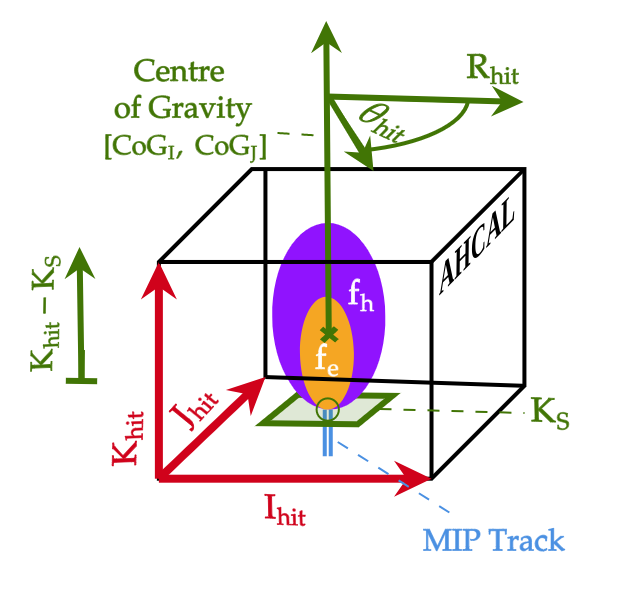}

\caption{Diagram illustrating the modified event coordinate system. The red axes indicate the readout of the event from the cells. The green axes indicate the transformed spatial co-ordinates, as defined from the lateral centre-of-gravity, indicated by $\text{CoG}_{I}$ and $\text{CoG}_{J}$ and the shower starting position, indicated by the green plane labelled $K_{S}$. }
\label{fig:Physics_RKS}
\end{figure}

The event coordinate system is changed for this study to reflect the shower development. The energy-weighted mean spatial position of the hadron shower in spatial coordinates is defined as a vector called 'centre-of-gravity' ($\mathrm{CoG} = \left[\mathrm{CoG}_{I}, \mathrm{CoG}_{J}, \mathrm{CoG}_{K}\right]$). Shower coordinates are converted to a cylindrical coordinate system, relative to the shower development axis, approximated by $\mathrm{CoG}_{I}$ and $\mathrm{CoG}_{J}$, and the shower starting depth, $K_{S}$. These coordinates act as the origin and are all measured in cell units. $K_{S}$ is calculated using a algorithm described in Ref.~\cite{AHCALKS}. The transformation results in three new spatial coordinates: a hit radius, $R_{\mathrm{hit}}  = \sqrt{(I_{\mathrm{hit}} - \mathrm{CoG}_{I})^2 + (J_{\mathrm{hit}} - \mathrm{CoG}_{J})^2}$, measured in cell units; a hit azimuthal angle, $\theta_{\mathrm{hit}} = \mathrm{arctan2}\left(J_{\mathrm{hit}} - \mathrm{CoG}_{J}, I_{\mathrm{hit}} - \mathrm{CoG}_{I} \right)$, where $\mathrm{arctan2}$ is the 2-argument arctangent, measured in radians, and a shower-start normalised depth, $K_{\mathrm{hit}} - K_{S}$, measured in layer indices. This coordinate system is advantageous to an SC algorithm because the hadron shower is represented independently of the lateral position of the hadron shower and the depth at which the hadron shower starts, and performs well for square cells of equal transverse size as in the AHCAL. Furthermore, these 'natural' spatial coordinates describe the lateral and longitudinal development of hadron showers more effectively than the raw event readout coordinates. They can be readily obtained from a well-separated single hadron shower. A visual representation of these coordinates is shown in Fig.~\ref{fig:Physics_RKS} for reference.

\subsection{Neural Network SC Method}\label{sec:NeutralNetworkModel}

SC models are typically trained indirectly since the $\frac{e}{h}$ fraction is unknown in a hadron shower event a priori. The resolution of a hadron calorimeter is described according to Eq.~\ref{eq:Resolution}:
\begin{equation}
    R = \frac{\sigma_{E}}{{E}} = \frac{a \cdot \unit{\sqrt{\qty{1}{\giga \electronvolt}}}}{\sqrt{E_{\mathrm{particle}}} } \oplus b,
    \label{eq:Resolution}
\end{equation}

where  $E$ and $\sigma_{E}$ are the mean and standard deviation of the response of the calorimeter to a hadron of $E_{\mathrm{particle}}$, $a$ describes the combined sampling and stochastic fluctuations experienced by the calorimeter, $b$ the quality of detector calibration, non-uniformities in signal collection, imperfections in calorimeter construction etc. and $\oplus$ addition in quadrature. This equation is valid under the assumption of a normally distributed response (i.e. full shower containment, negligible electronics noise). Since reductions in $\sigma_{E}$ imply compensation due to a smaller $a$, $\chi^{2}$ minimisation of the calorimeter response to the known particle energy may be used to optimise SC algorithms. 
However, the lack of available high-statistics training samples for SC at finely binned particle energies tends to result in undesirable network biases that limit the algorithms' general applicability. In particular, two failure modes have been observed \cite{erik}: the 'classification' of the hadron showers by calorimeter response and bias to the training sample's upper and lower particle energy bins. 


In particular, two failure modes have been observed \cite{erik}: the 'classification' of the hadron showers by calorimeter response and bias to the training sample's upper and lower particle energy bins.

\begin{figure*}
    \centering
    \includegraphics[width=\linewidth]{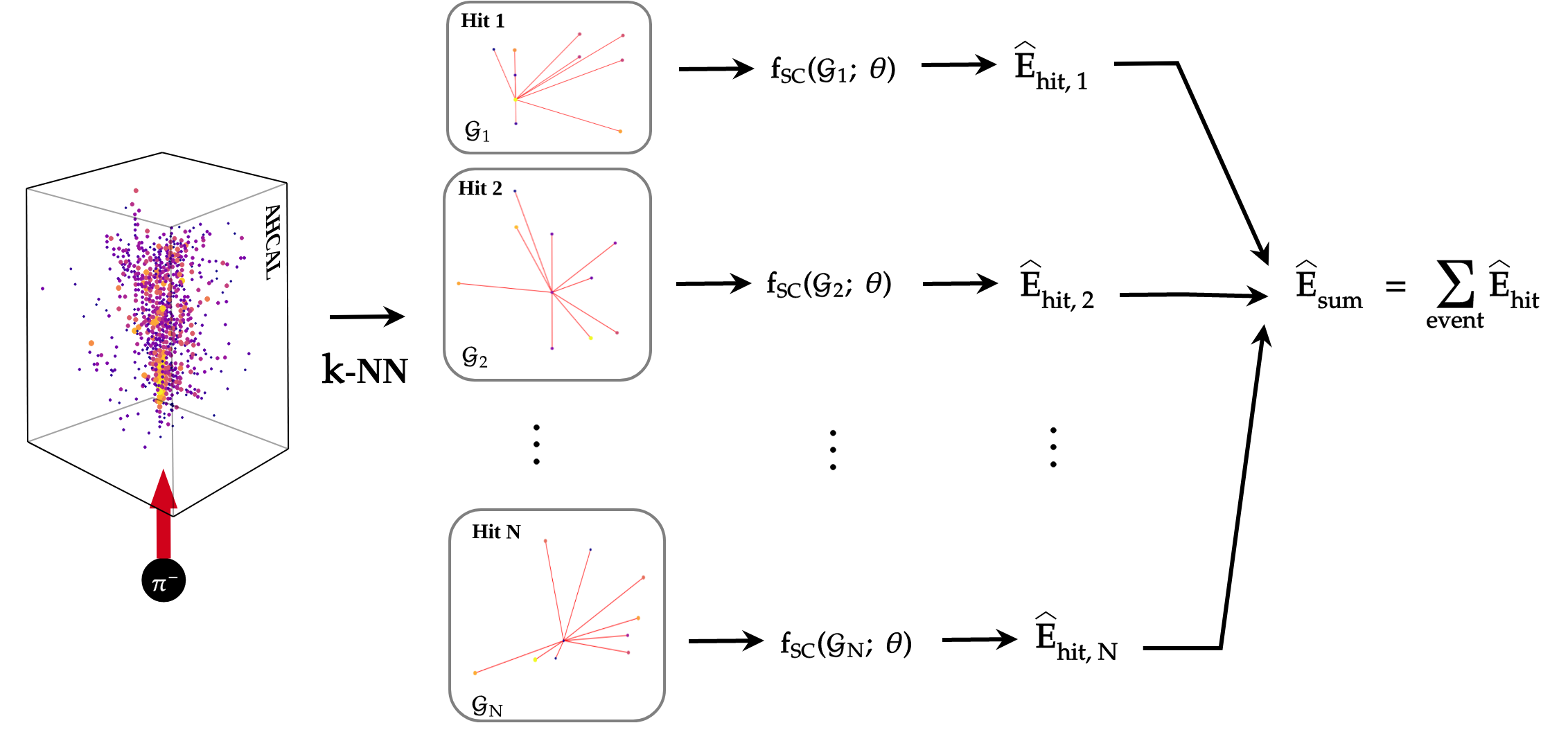}
    \caption[Illustration of Proposal For Reducing Bias in SC Training by Model Design]{Illustration of a method for software compensation by which biasing may be reduced. First, a hadron shower of $N$ measured with AHCAL, indicated by the event display on the left, is decomposed into a series of $k$-NN cluster graphs, indicated by the vertices and red lines, indicating edges between them, denoted $\mathcal{G}$. At this stage, each active cell is now represented as a local neighbourhood graph, $\mathcal{G}_{i}$, where $i$ is the index of the active cell. This diagram shows the case of 9 nearest neighbours for illustration. Next, for each cluster, a SC model, $f_{\mathrm{SC}}(\mathcal{G}_{i}; \theta)$, is applied to each graph, where $\theta$ is the vector of the model's free parameters, producing an attenuated or enhanced calorimeter response to the kernel cell of $\mathcal{G}_{i}$, $\widehat{E}_{\mathrm{hit}, i}$. The sum of the individually weighted active cells is then the compensated calorimeter response, $\widehat{E}_{\mathrm{sum}}$.}
    \label{fig:ModelIdea}
\end{figure*}

A neural network was, therefore, designed to overcome the limitations of energy biasing. The proposed model was designed to use $k$ nearest-neighbour ($k$-NN) clustering in the event coordinates defined in Sec.~\ref{sec:CALICE_AHCAL} to obtain a local estimate of the energy density in space and time. A $k$-NN cluster consists of the $k$ nearest points in the event in space, energy and, optionally, time in terms of the square Euclidean distance, $\mathrm{d}s^2 = \mathrm{d}R_{\mathrm{hit}}^{2} + \mathrm{d}\theta_{\mathrm{hit}}^{2} + \mathrm{d}(K_{\mathrm{hit}} - K_{\mathrm{S}})^{2} + \mathrm{d}\log{E_{\mathrm{hit}}}^{2} + \mathrm{d}\operatorname{arcsinh}{T_{\mathrm{hit}}}^{2}$, where $R_{\mathrm{hit}}$ and $K_{\mathrm{hit}} - K_{\mathrm{S}}$ are in units of cells, $\theta_{\mathrm{hit}}$ is in units of radians, $E_{\mathrm{hit}}$ is in calibrated MIP units and $T_{\mathrm{hit}}$ is in units of nanoseconds. $E_{\mathrm{hit}}$ and $T_{\mathrm{hit}}$ are transformed to reduce the skewness of these variables. In practice,  $k$-NN clustering is implemented by calculating the negative square distance matrix $-D^{2}_{ij} = -|x_{i}|^{2}- |x_{j}|^{2} + 2\cdot \left\langle x_{i}, x_{j} \right\rangle$ where $x_{i}$ and $x_{j}$ are co-ordinates of individual hits in space and time with indices $i$ and $j$,  $|x_{i,j}|^{2}$ is the absolute square of $x_{i}$ or $x_{j}$, and  $\left\langle x_{i}, x_{j} \right\rangle$ is the inner product of $x_{i}$ and $x_{j}$. The columns of the $D$ matrix are ranked based on their proximity to zero. The top $k$ elements are then selected for each column, representing the $k$-nearest neighbours for each data point or coordinate, and the vectors between the seed cell and the other cells of the cluster are calculated. Each cluster is then treated independently by the neural network. With context to highly-granular calorimetry, this gives the local energy density surrounding a particular active cell during an event. The neural network was designed based on a single EdgeConv operator, introduced in the DGCNN graph neural network model \cite{DGCNN}. The value of $k$ was optimised using a hyperparameter scan to 20 cells using the Optuna hyperparameter optimisation tool \cite{Optuna}, as shown in Table \ref{tab:hyperparameters}.
The most critical aspect of the neural network design is that each cluster is operated independently of all others. This choice means that the capacity for the neural network to learn biased features of the training data, such as overall shower shape and energy, can be reduced compared to the case where the entire shower is presented as an input.  This is because the network is subjected only to the local distributions of individual clusters of active cells. Therefore, the neural network is guided to infer the appropriate attenuation or enhancement of the calorimeter response from the energy distribution local to each active cell. The idea is summarised in Fig.~\ref{fig:ModelIdea}.

The neural network architecture consists of five main stages:

\begin{itemize}

    \item \textit{Input}: The neural network is provided two inputs. The first is the hadron shower event in natural/transformed coordinates ($\left[R_{\mathrm{hit}}, \theta_{\mathrm{hit}}, K_{\mathrm{hit}}-K_{S}, \log{E_{\mathrm{hit}}}, \operatorname{arcsinh }{t_{\mathrm{hit}}}\right]$), where $\operatorname{arcsinh}{t_{\mathrm{hit}}}$ is optional. The second is the original cell energy, $E_{\mathrm{hit}}$, which is used to inform the neural network of the output scale of the compensated energy;

    \item \textit{$k$-NN clustering}: As a pre-processing step, the neural network clusters the input according to the $k$-nearest neighbours. Their positions and vectors to their positions are calculated;
    
    \item \textit{Addition of Dimensions}: Dimensions are added to each cluster using a module consisting of three sequential 2D fully connected convolutional layers 12, 24, and 48 channels, each using leaky ReLU activation and instance normalisation. Each new dimension is calculated using information from the inputs. 
    
    \item \textit{Processing}: Each cluster is passed through a deep processing layer consisting of 3 sequential 2D fully connected layers of 48 channels, each using leaky ReLU activation and instance normalisation;
    
    \item \textit{Aggregation}: The maximum, mean and variance of the cluster dimension $k$ are used as activation values for the cluster. These are concatenated with the cell energies of the event for each active cell;
    
    \item \textit{Output}: The final layers of the network are five dense layers, with 1024, 512, 256 and 128 channels and leaky ReLU activation, with an output layer with ReLU activation such that the final output is positive. All dense layers, excluding the final layer, include dropout with probability $p_{\mathrm{dropout}}$. The neural network's final output is a single value for each active cell: the compensated hit energy, $\widehat{E}_{\mathrm{hit}}$. The sum of these outputs yields the total compensated response, $\widehat{E}_{\mathrm{sum}} = \sum^{\mathrm{event}}
    \widehat{E}_{\mathrm{hit}}$, where $\widehat{E}_{\mathrm{hit}} \in \left[0, \infty \right]$ is the compensated cell energy.
    
\end{itemize}

\begin{figure*}
    \centering
    \includegraphics[width=\linewidth]{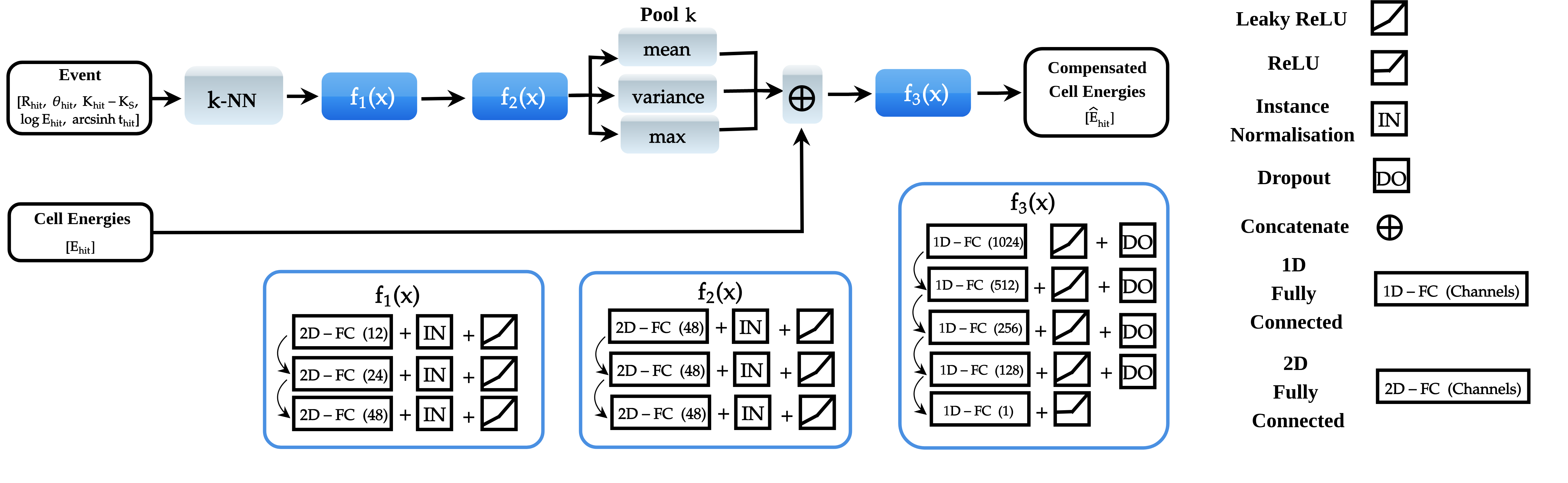}
    \caption{Flowchart describing the proposed neural network for software compensation studied in this paper. The black, blue and grey colour coding indicates inputs and outputs, convolutional operations and general operations, respectively. Additional operations are specified on the right of the figure.}
    \label{fig:SCNet}
\end{figure*}

A diagram representing the proposed neural network architecture is shown in Fig.~\ref{fig:SCNet}.

\subsection{Control SC Method}\label{sec:ControlMethod}

 The neural network was compared to the standard CALICE software compensation method called \textit{'local software compensation'}, abbreviated hereafter as the control method and based on \cite{LSC}, is described as follows.

The $E_{\mathrm{hit}}$ distribution is binned in deciles (i.e. a 10\% probability for a given $E_{\mathrm{hit}}$ to be found in any one of the bins). For each bin, an appropriate function is used for weighting. A function approximator in the form of a second-order Chebyshev polynomial of the first kind, $\omega_{b}$, is defined as a function of the total calorimeter response, $E_{\mathrm{sum}}$, scaled using a factor, $S$, such that $E_{\mathrm{sum}}/S \in \left[0, 1\right]$ for the typical range of hadron shower energies of AHCAL ($S = \qty{150}{\giga \electronvolt}$). $\omega_{b}$ has three free parameters, $\alpha_{\mathrm{b}}$, $\beta_{\mathrm{b}}$ and $\gamma_{\mathrm{b}}$, shown in Eq.~ \ref{eq:LSC_cheby}:

\begin{equation}
    \omega_{b}(E_{\mathrm{sum}};  S, \alpha_{\mathrm{b}}, \beta_{\mathrm{b}}, \gamma_{\mathrm{b}}) =  \alpha_{\mathrm{b}}+\beta_{\mathrm{b}} \cdot \left(\frac{E_{\mathrm{sum}}}
{S}\right)+\gamma_{\mathrm{b}} \cdot \left(2\left(\frac{E_{\mathrm{sum}}}{S}\right)^{2}-1\right)
\label{eq:LSC_cheby}
\end{equation}

For each bin, the corresponding weight is calculated. Finally, the energy of each active cell within the ranges defined by bin $b$ is scaled by $\omega_{\mathrm{b}}$.

\begin{equation}
    \widehat{E}_{\mathrm{sum}} = \sum^{\mathrm{bins}}_{\mathrm{b}} \omega_{\mathrm{b}} \cdot E_{\mathrm{sum}, \mathrm{b}}
\end{equation}

\begin{figure}
    \centering
    \includegraphics[width=0.6\textwidth]{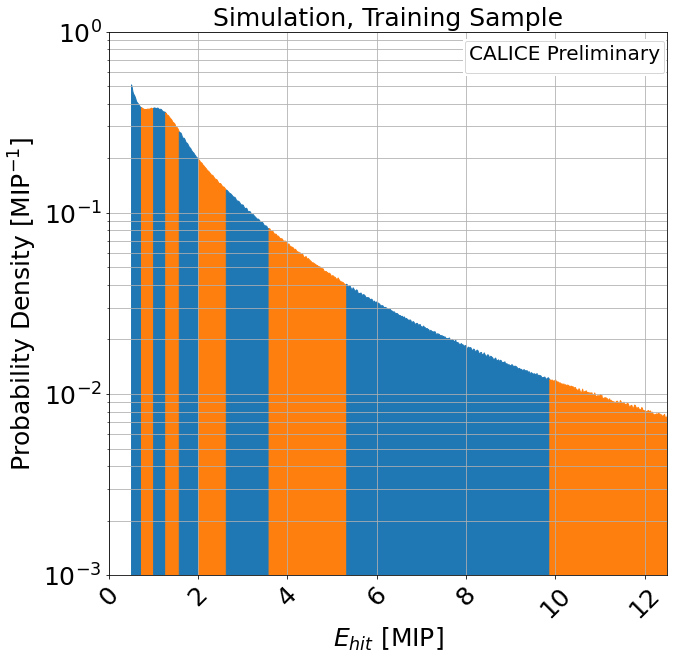}
    \caption{Histogram showing the ten decile bin ranges of the $E_{\mathrm{hit}}$ distribution, shown in alternating blue and orange, are shown for the training sample. Each bin corresponds to a total cumulative probability of 10\% (deciles). The bin ranges extracted from the training sample discussed in Section \ref{sec:Datasets}  are shown in Appendix Table \ref{tab:LSCWeights}.}
    \label{fig:SCNet_LSCBins}
\end{figure}

The idea underlying this method is that higher hit energy bins attenuate the energy, as these are more likely to belong to an EM fraction and enhance the energy of low energy bins, which are more likely to belong to the HAD fraction. An example of the ten bins selected for the study is in Fig.~\ref{fig:SCNet_LSCBins}.

\subsection{Datasets and Training}\label{sec:Datasets}

Both the neural network model defined in Section \ref{sec:NeutralNetworkModel} and the control model defined in Section \ref{sec:ControlMethod} were trained and validated using experimental data from a CALICE test beam study at the Super Proton Synchrotron at CERN in 2018, as well as a simulated dataset thereof \cite{Testbeam}. Each case was studied separately. Both simulation and experimental data were used for training and evaluation, respectively. The showers were produced from $\pi^{-}$ hadron shower events observed with the AHCAL detector. The simulation of the particle showers was achieved using \texttt{Geant4} \cite{Geant4}, with a full detector simulation developed using \texttt{DD4hep} \cite{DD4Hep}. Additional effects, such as digitisation of the analogue signal and reconstruction of the detector variables, were achieved for both simulation and data using \texttt{CALICESoft} \cite{CALICESoft}. Timing information from experimental data is not studied due to comparatively poor timing resolution arising from chip occupancy effects \cite{Lorenz}. A MIP-to-GeV calibration factor of \qty{37.3}{\text{MIP} \per \giga \electronvolt} was used \cite{Olin}. The statistics of the training, validation and test datasets are shown in Table \ref{tab:SCNet_EventTable}.

The following selection criteria were applied: 

\begin{itemize}

    \item events were required to be identified using the standard CALICE particle identification algorithm \cite{PID} as being a single particle and having less than a \qty{0.5}{\percent} probability of being a muon to exclude non-showering, 'punch-through' pions;
    
    \item the 38$^{\mathrm{th}}$ layer of the AHCAL was ganged and required special treatment beyond the scope of this paper.  Therefore, energy deposits were considered up to the 38$^{\mathrm{th}}$ layer of the calorimeter;
    
    \item events were selected to have a track position with a corresponding position inside the $24\times\qty{24}{\text{ cell}}$ AHCAL front-face and a shower starting layer within layers $1$-$4$ of the AHCAL calorimeter. This choice was made to reduce the effect of longitudinal and lateral leakage on the experiment. These cuts were supplemented by an additional cut using the TCMT detector to only measure detector resolution and linearity. This criterion requires the TCMT to measure a total deposited energy of less than $\qty{25}{\text{MIP}}$ ($E^{\text{TCMT}}_{\text{sum}}<\qty{25}{\text{MIP}}$).
    
\end{itemize}

\begin{table*}
\centering
\caption[Table of Events Used For Training and Evaluating SC Models]{Table of events used for training SC models after all cuts except the TCMT cut (shown separately), split into simulation and data and by the testing, training and validation samples and by data and simulation. Hyphens indicate 0 events. }

\label{tab:SCNet_EventTable}
\vspace{0.2in}
\scriptsize{
\begin{tabular}{lllllllll}
\toprule
Type & \multicolumn{4}{l}{June 2018 SPS Testbeam Data} & \multicolumn{4}{l}{Simulation} \\
Sample &       Test & Test & Training & Validation &   Test & Test & Training & Validation \\
 &        &  + TCMT Cut &  &  &   &  + TCMT Cut &  &  \\
$E_{\mathrm{particle}}$ [\unit{\giga \electronvolt}] &               &          &            &            &          &            \\
\midrule
10             &          6472 & 6460 &   51773 &       6472 &       20826 & 20759 &    18719 &       2080 \\
15             &             - &  -   &      - &          - &      21969 & 21685 &       - &          - \\
20             &          9439 & 9233  &  75512 &       9439 &      23425 & 22808 &    21428 &       2381 \\
25             &             - &  - &      - &          - &      25193 & 24124 &        - &          - \\
30             &             - &  - &      - &          - &      24031 & 22491 &    21901 &       2434 \\
35             &             - &  - &      - &          - &      24154 & 22065 &        - &          - \\
40             &         10384 &  9378 &  83064 &      10383 &      24195 & 21513 &    23552 &       2617 \\
45             &             - &  - &      - &          - &      23122 & 19981 &         - &          - \\
50             &             - &  - &      - &          - &      27337 & 22889 &    24737 &       2749 \\
55             &             - &  - &      - &          - &      19636 &    16009 &    - &          - \\
60             &         13223 &  10684 & 105782 &      13223 &      22503 & 17728 &    24479 &       2720 \\
65             &             - &  - &      - &          - &      25584 & 19374 &        - &          - \\
70             &             - &  - &      - &          - &      18951 & 13889 &    24864 &       2763 \\
75             &             - &  - &      - &          - &      15827 & 11204 &        - &          - \\
80             &         11666 &  8298 &  93325 &      11666 &      22272 & 15165 &    25308 &       2813 \\
85             &             - &  -  &      - &          - &      22577 & 14875 &        - &          - \\
90             &             - &  - &      - &          - &      26210 & 16618 &        - &          - \\
95             &             - &  - &      - &          - &      20605 & 12475 &        - &          - \\
100            &             - &  - &      - &          - &      17706 & 10385 &        - &          - \\
105            &             - &  - &      - &          - &      17410 & 9873 &        - &          - \\
110            &             - &  - &      - &          - &      16885 & 9161 &        - &          - \\
115            &             - &  - &      - &          - &      18706 & 9820 &        - &          - \\
120            &         10713 &  5829 &    85701 &      10713 &      18192 & 9239        - &          - \\

\bottomrule
Total Events          &         61897 & 49882 &   495157 &      61896 &   497316 & 395131 &  184988 &      20557 \\
\bottomrule
\end{tabular}
}

\end{table*}

The cuts applied, not including the TCMT cut, remove around two-thirds of the original sample. For the measurement of resolution, Eq. \ref{eq:Resolution} cannot be used if the AHCAL experiences longitudinal shower leakage since the distributions exhibit a skewed 'leakage tail' that deviates from the expected Gaussian response distribution. To resolve this, the TCMT is employed to tag and cut events likely to have a fraction of leakage energy.

\begin{figure*}[htb!]
    \centering
    \subfloat[\label{fig:SCNet_CheckPlot_ESum10TCMT_Sim}]{
    \centering
    \includegraphics[width=0.44\linewidth]{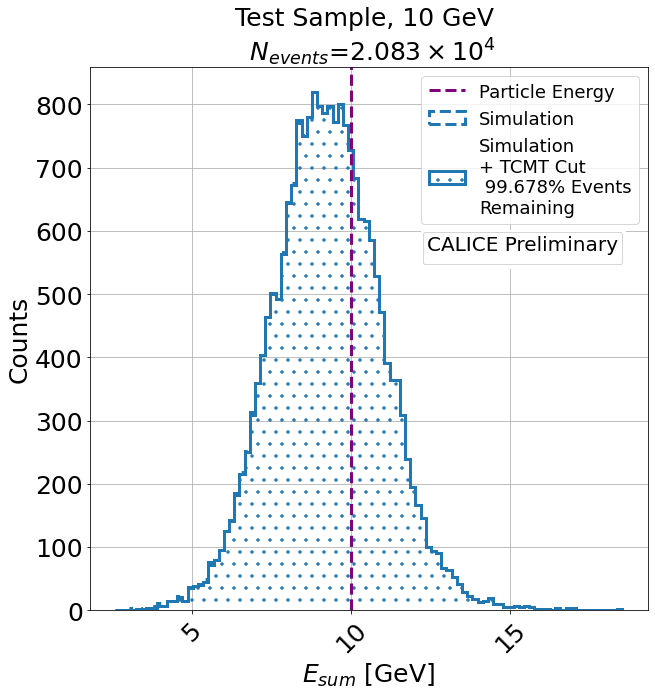}
    }
    \hfill
    \subfloat[\label{fig:SCNet_CheckPlot_ESum80TCMT_Sim}]{
    \centering
    \includegraphics[width=0.44\linewidth]{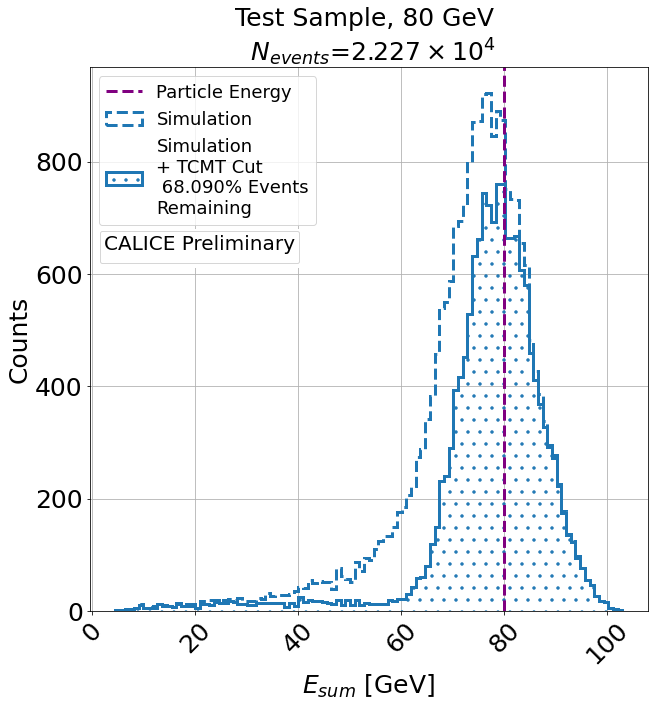}
    }
    \vfill
    \subfloat[\label{fig:SCNet_CheckPlot_ESum10TCMT_SimDataComp}]{
    \centering
    \includegraphics[width=0.44\linewidth]{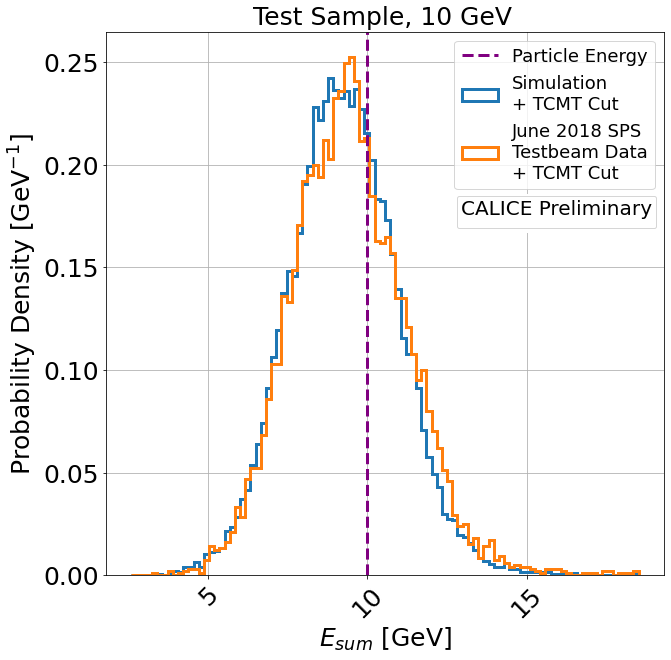}
    }
    \hfill
    \subfloat[\label{fig:SCNet_CheckPlot_ESum80TCMT_SimDataComp}]{
    \centering
    \includegraphics[width=0.44\linewidth]{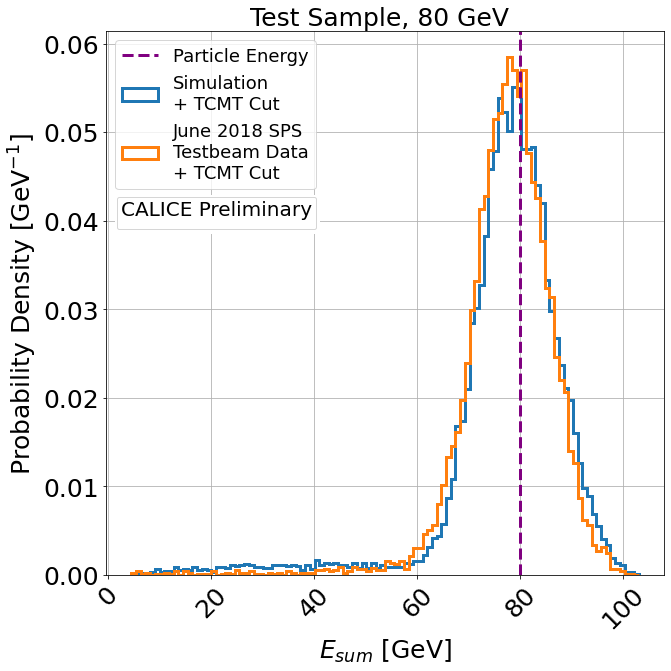}
    }
    
    \caption{ Fig.~\ref{fig:SCNet_CheckPlot_ESum10TCMT_Sim} and Fig.~\ref{fig:SCNet_CheckPlot_ESum80TCMT_Sim} show the reconstructed AHCAL energy distributions of simulation for \qty{10}{\giga \electronvolt} and \qty{80}{\giga \electronvolt} $\pi^{-}$ hadron showers. The dashed line and the solid line filled with dots indicate the distribution before and after the applied TCMT cut. Fig.~\ref{fig:SCNet_CheckPlot_ESum10TCMT_SimDataComp} and Fig.~\ref{fig:SCNet_CheckPlot_ESum80TCMT_SimDataComp} show response distributions of the test samples of simulation and June 2018 Testbeam data before and after the cut was applied, for \qty{10}{\giga \electronvolt} and \qty{80}{\giga \electronvolt} $\pi^{-}$ hadron showers. }
    \label{fig:SCNet_CheckPlot_TCMT}

\end{figure*}

Examples of the effect of the applied TCMT cut are shown in Fig.~\ref{fig:SCNet_CheckPlot_TCMT}. Fig.~\ref{fig:SCNet_CheckPlot_ESum10TCMT_Sim} indicate the cut has practically no effect on \qty{10}{\giga \electronvolt} hadron showers, while Fig.~\ref{fig:SCNet_CheckPlot_ESum80TCMT_Sim} show that the cut significantly reduces the leakage tail of the response distribution of \qty{80}{\giga \electronvolt} hadron showers, resulting in a more Gaussian distribution at energies where leakage is observed. This means that the cut can be used to evaluate the resolution of the AHCAL alone. Fig.~\ref{fig:SCNet_CheckPlot_ESum10TCMT_Sim} and Fig.~\ref{fig:SCNet_CheckPlot_ESum80TCMT_Sim} indicate that the energy distribution of simulation is similar to experimental data and that the cut has a similar effect on both. This means the cut can be applied without modification in both cases.

As a caveat, this method is sub-optimal based on the considerable longitudinal leakage experienced by the AHCAL and will bias the resolution measurement to more compact showers. Therefore, example distributions illustrating the performance of the software compensation without the TCMT cut are supplied for reference in the results shown in Section \ref{sec:Results}, to illustrate the performance of each method without this bias.

The training and validation dataset consisted of simulated showers induced by $\pi^{-}$ hadrons with $E_{\mathrm{particle}}$ in the range $10$-$\qty{80}{\giga \electronvolt}$, in increasing steps of $\qty{10}{\giga \electronvolt}$. By contrast, the test sample contained showers induced by $\pi^{-}$ hadrons with $E_{\mathrm{particle}}$ in the range $10$-$\qty{120}{\giga \electronvolt}$, in increasing steps of $\qty{5}{\giga \electronvolt}$. The finer granularity tests the hypothesis that the neural network is unbiased to the particular particle energies used for training. Energies higher than the training range are included to test the generalisation capacity of each compensation method. The whole range of energies is used for training and testing for experimental data.

Two independent networks based on the model defined in Section \ref{sec:NeutralNetworkModel} were trained on the training dataset: one without timing information and one with timing information. The proposed compensation networks were developed in \texttt{PyTorch} \cite{PyTorch} and trained using the \texttt{PyTorch Lightning} research framework \cite{PyTorchLightning} on an NVidia V100 GPU. The ADAM optimiser was used to improve the convergence rate for ten epochs, with early stopping applied.  The hyperparameters used for training are shown in Table \ref{tab:hyperparameters}. Hyperparameter optimisation was achieved by optimising the networks many times with different starting parameters for $k$, learning rate, and dropout probability. Varying $\beta_{1}$ and $\beta_{2}$ resulted in large fluctuations in performance and were therefore held at nominal values. The hyperparameter search program \texttt{Optuna} optimised the hyperparameters for 50 trials, with 10 epochs per trial. To speed up convergence, trials were rejected using ‘median pruning’, where a trial is pruned if its best intermediate result is worse than the median of intermediate results of previous trials at the same epoch. 

\begin{table}[]
    \caption{Table of hyperparameters used to train the neural network. In this table, $\beta_{1}$ and $\beta_{2}$ are the ADAM momentum parameters, $p_{\mathrm{dropout}}$ is the dropout probability, and $k$ is the number of nearest-neighbours per cluster. The parameters were informed by a hyperparameter scan using \texttt{Optuna} \cite{Optuna}.}
    \label{tab:hyperparameters}
    \vspace{0.2in}

    \centering
    \small{
    \begin{tabular}{c|c}
        \hline
        Parameter & Value  \\
        \hline
        Learning Rate &  $9\times 10^{-5}$ \\
        Batch Size & $32$ \\
        $\beta_{1}$ & $0.9$ \\
        $\beta_{2}$ & $0.999$ \\
        $p_{\mathrm{dropout}}$ & $0.15$ \\
        $k$ & 20 \\
        \hline
    \end{tabular}
    }

\end{table}

 The control method was also trained using the training dataset, using the \texttt{MIGRAD} algorithm of the \texttt{Minuit} minimisation program \cite{Minuit}. Weights were initialised such that the compensation algorithm acted as the identity operator ($\alpha_{\mathrm{b}}=1$, $\beta_{\mathrm{b}}=0$, $\gamma_{\mathrm{b}}=0$). 

 The loss was chosen to be the $\chi^{2}$ goodness-of-fit of the compensated energy to the known particle energy of the hadron shower:

\begin{equation}
    \mathcal{L}(\widehat{E}_{\mathrm{sum}}; E_{\mathrm{particle}}) = \frac{\left(\widehat{E}_{\mathrm{sum}} - E_{\mathrm{particle}}\right)^{2}}{E_{\mathrm{particle}}\cdot(\qty{1}{\giga \electronvolt})}
\end{equation}

The denominator in the loss arises from the uncertainty on the Poisson-distributed sampling quanta measured by the calorimeter, $\sigma_{E} = a\cdot\sqrt{E_{\mathrm{particle}}}$. The dummy constant of \qty{1}{\giga \electronvolt} in the denominator is formally included to make the loss unitless and merely acts to scale the loss. The mean loss was used for both implementations to optimise the control and network methods. For the network methods, the epoch with the smallest mean loss of the validation sample was chosen for further study. The control method was minimised with the MIGRAD algorithm until the mean training loss reached convergence. 

The training process involved utilising both simulation and data training samples from Table \ref{tab:SCNet_EventTable} to train separate models for both cases, which were subsequently evaluated on their respective test samples. Consequently, the model's performance was assessed independently for scenarios in which it was trained with simulation or experimental data.

\section{Results}\label{sec:Results}

Each trained model was applied to the test sample. The effect of compensation was then analysed for each method. 

\subsection{Example Response Distributions}

\begin{figure*}[t]

\subfloat[%
 \label{fig:10GeV}
]{\includegraphics[width=0.44\linewidth]{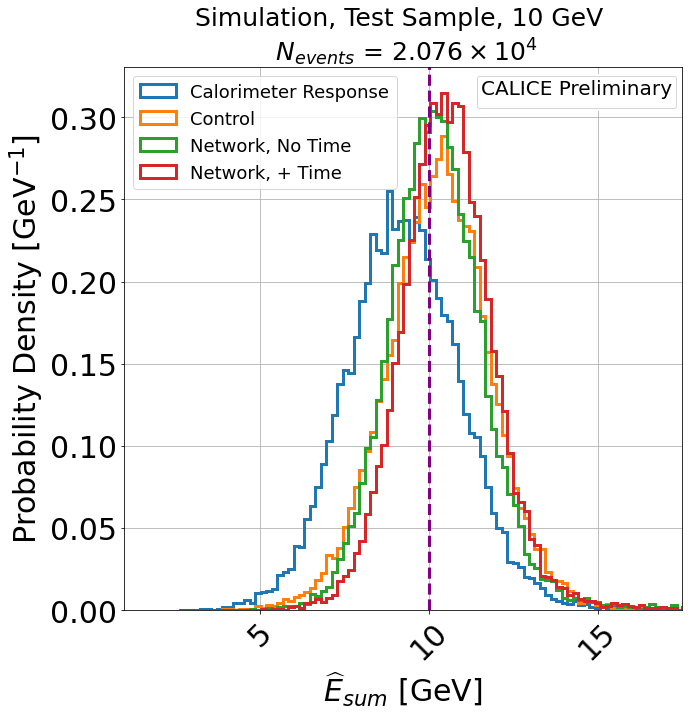}}
\hfill
\subfloat[%
 \label{fig:35GeV}
]{\includegraphics[width=0.44\linewidth]{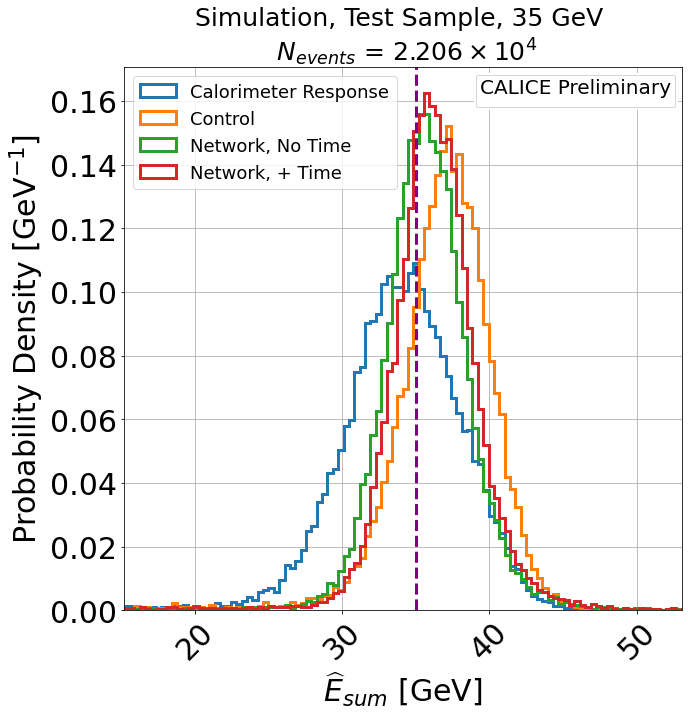}}
\hfill
\subfloat[%
 \label{fig:80GeV}
]{\includegraphics[width=0.44\linewidth]{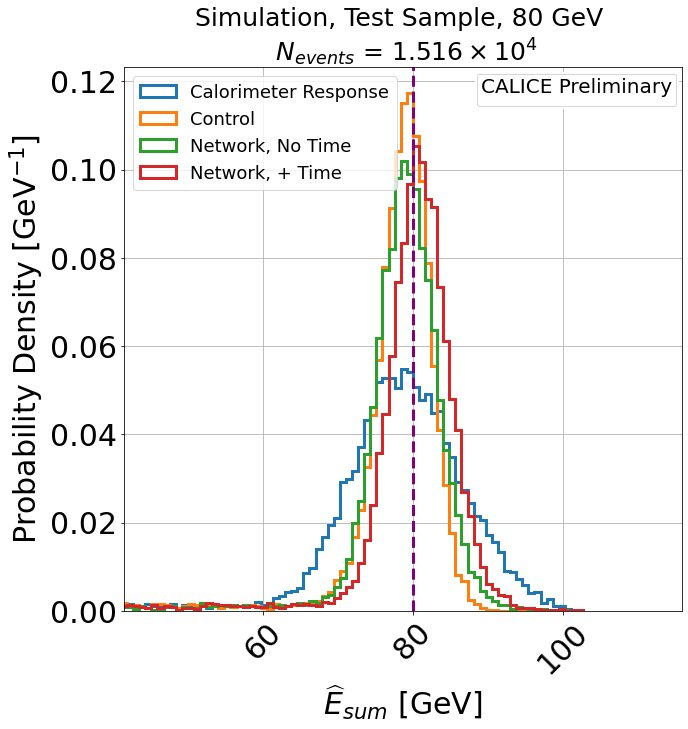}}
\hfill
\subfloat[%
 \label{fig:120GeV}
]{\includegraphics[width=0.44\linewidth]{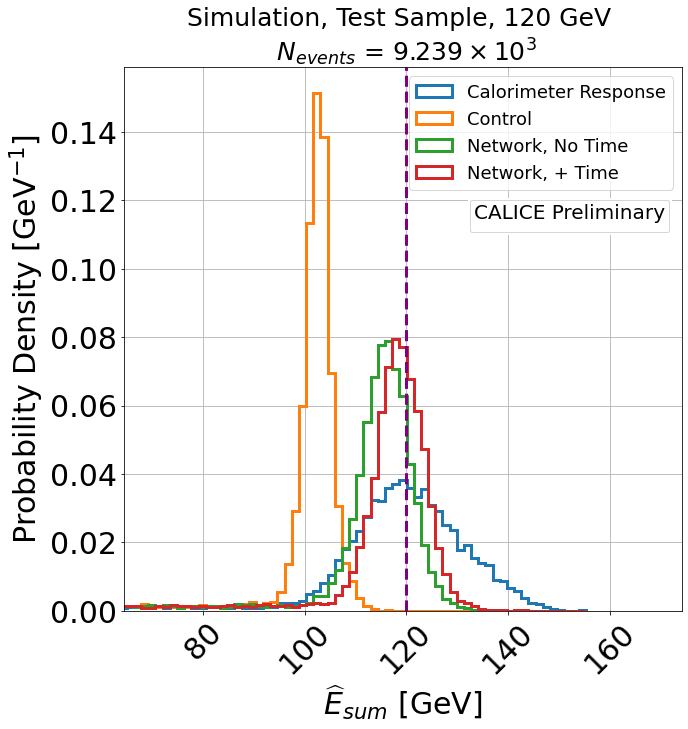}}

\caption{Example normalised histograms showing the calorimeter response before and after compensation applied to the simulated test dataset of Table \ref{tab:SCNet_EventTable}, with the TCMT cut applied. Samples of \qty{10}{\giga \electronvolt}, \qty{40}{\giga \electronvolt}, \qty{80}{\giga \electronvolt} and \qty{120}{\giga \electronvolt} hadron shower energies are shown. Blue lines indicate intrinsic calorimeter response, while orange, green and red lines indicate the control, network without and network with time compensation methods, respectively. $E_{\mathrm{particle}}$ is indicated as a dashed purple line. The number of events shown in the title indicate the corresponding sample sizes from Table \ref{tab:SCNet_EventTable}. }
\label{fig:Spectra_sim}
\end{figure*}

\begin{figure*}

\centering
\subfloat[%
 \label{fig:10GeV_Sim_NoTC}
]{\includegraphics[width=0.44\linewidth]{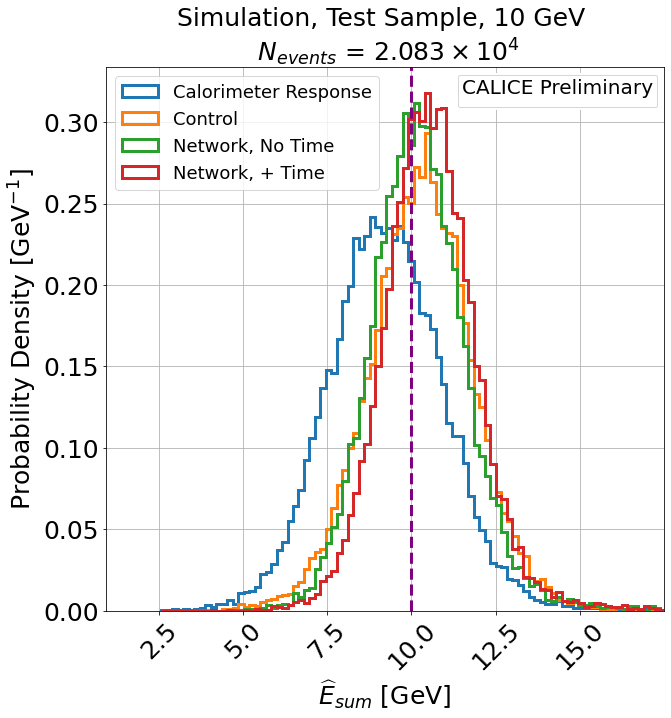}}
\hfill
\subfloat[%
 \label{fig:40GeV_Sim_NoTC}
]{\includegraphics[width=0.44\linewidth]{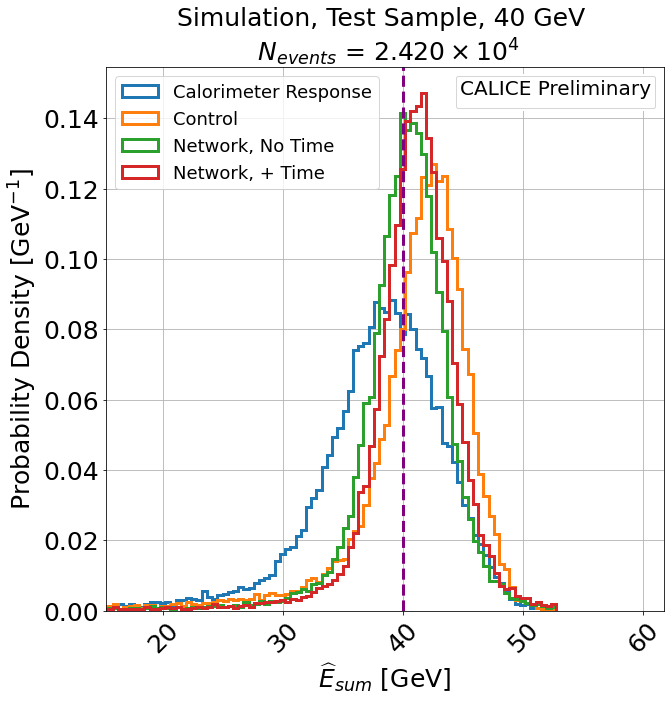}}
\vfill
\subfloat[%
 \label{fig:80GeV_Sim_NoTC}
]{\includegraphics[width=0.44\linewidth]{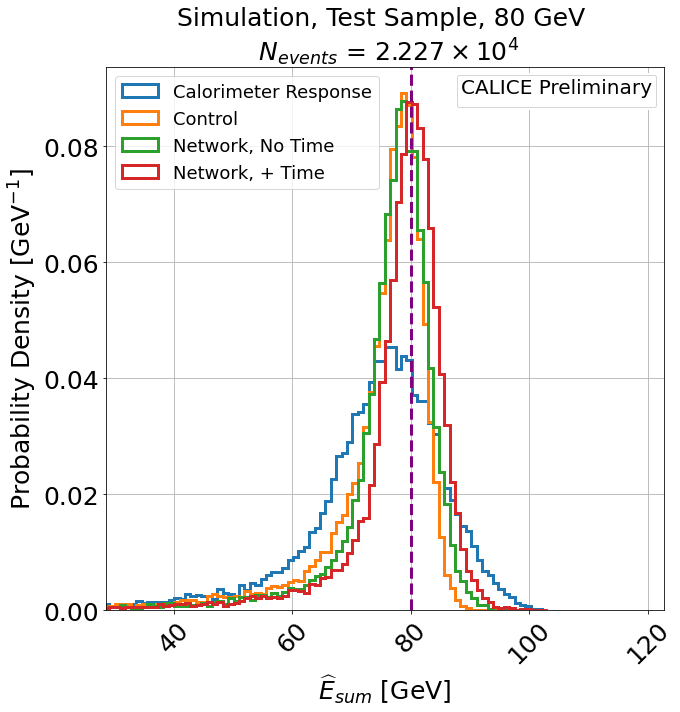}}
\hfill
\subfloat[%
 \label{fig:120GeV_Sim_NoTC}
]{\includegraphics[width=0.44\linewidth]{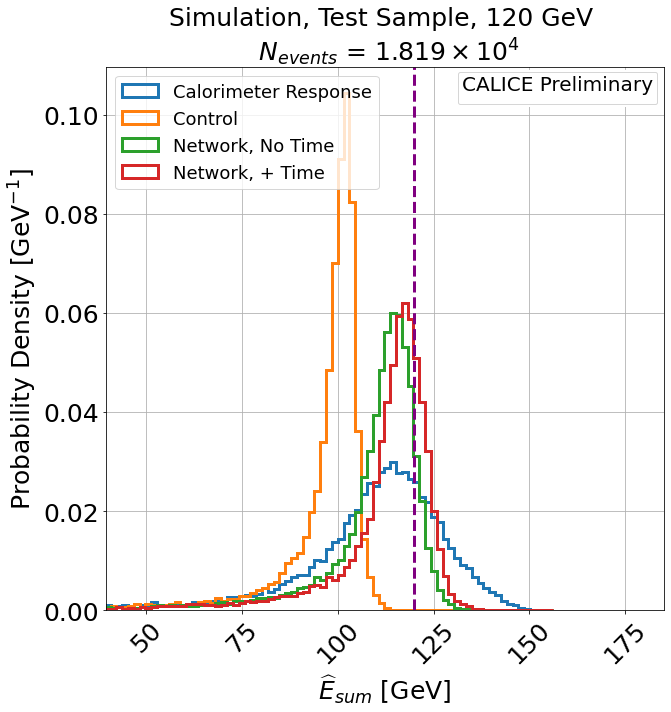}}

\caption{Example normalised histograms showing the calorimeter response before and after compensation applied to the simulated test dataset of Table \ref{tab:SCNet_EventTable}, without the TCMT cut applied. Else, as in Fig.~\ref{fig:Spectra_sim}.}

\label{fig:Spectra_Sim_NoTC}
\end{figure*}

\begin{figure*}

\subfloat[%
 \label{fig:10GeV_Data}
]{\includegraphics[width=0.44\linewidth]{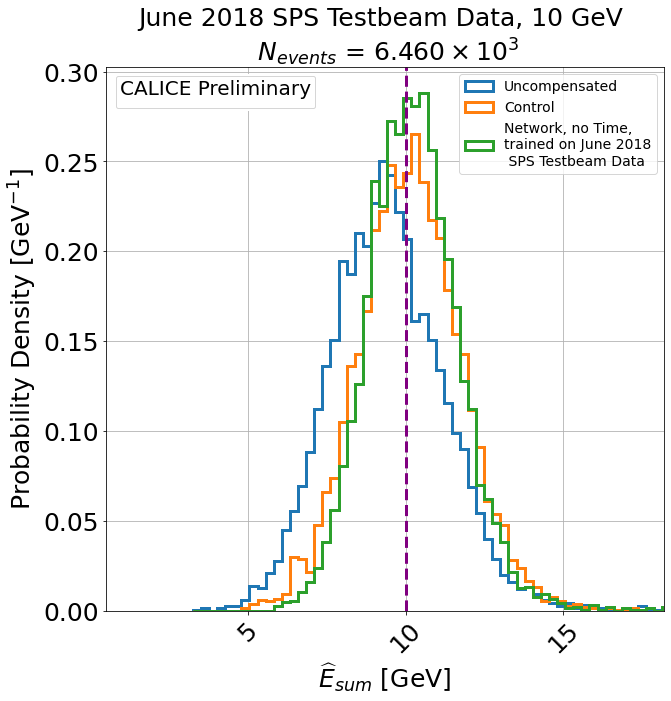}}
\hfill
\subfloat[%
 \label{fig:40GeV_Data}
]{\includegraphics[width=0.44\linewidth]{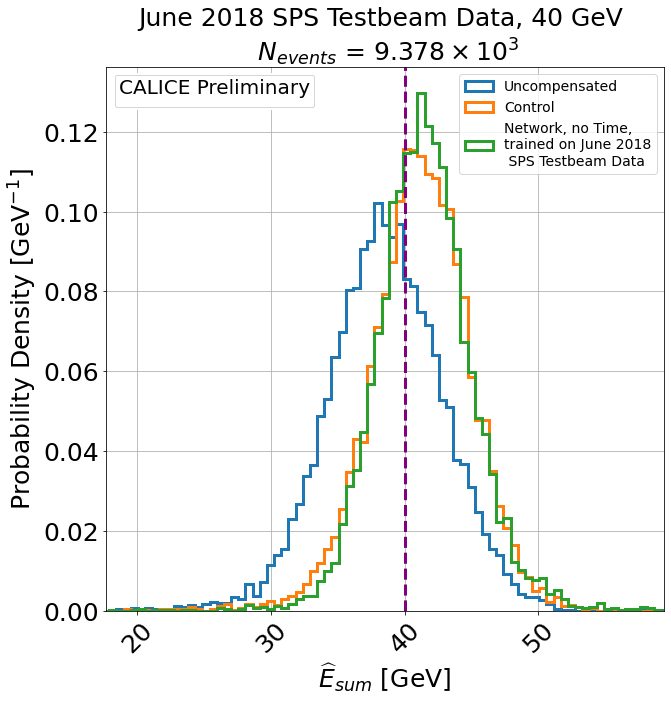}}
\hfill
\subfloat[%
 \label{fig:80GeV_Data}
]{\includegraphics[width=0.44\linewidth]{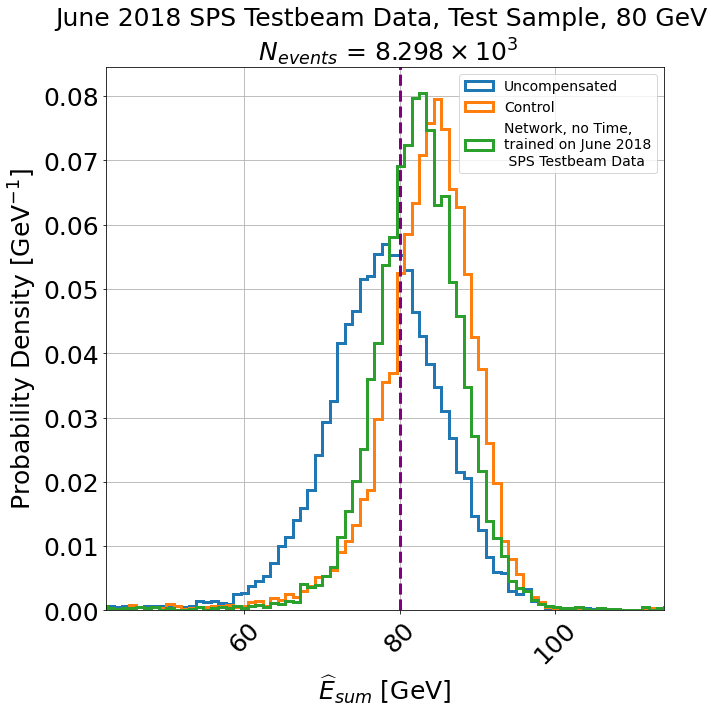}}
\hfill
\subfloat[%
 \label{fig:120GeV_Data}
]{\includegraphics[width=0.44\linewidth]{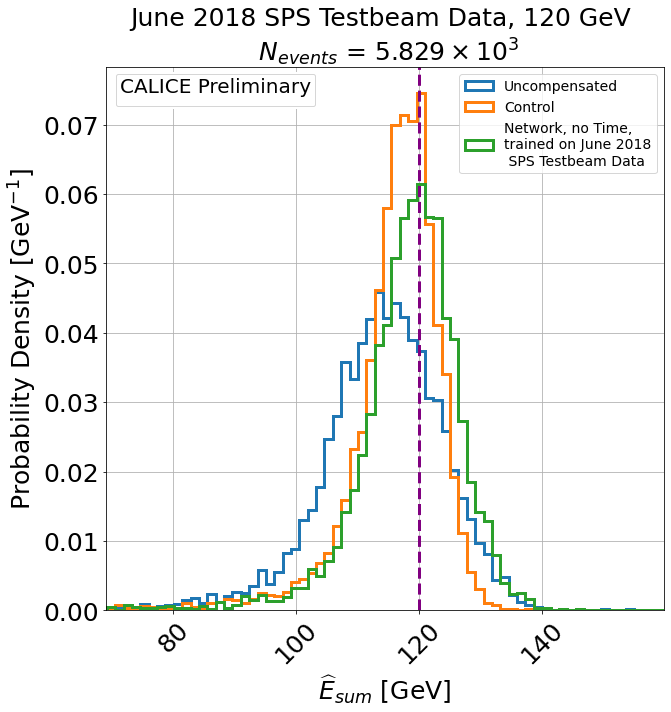}}

\caption{Example normalised histograms showing the calorimeter response before and after compensation applied to the 2018 June Testbeam test dataset of Table \ref{tab:SCNet_EventTable}, with the TCMT cut applied. Samples of \qty{10}{\giga \electronvolt}, \qty{40}{\giga \electronvolt}, \qty{80}{\giga \electronvolt} and \qty{120}{\giga \electronvolt} hadron shower energies are shown. Else, as in Fig.~\ref{fig:Spectra_sim}. }
\label{fig:Spectra_data}
\end{figure*}

\begin{figure*}

\centering
\subfloat[%
 \label{fig:10GeV_Data_NoTC}
]{\includegraphics[width=0.44\linewidth]{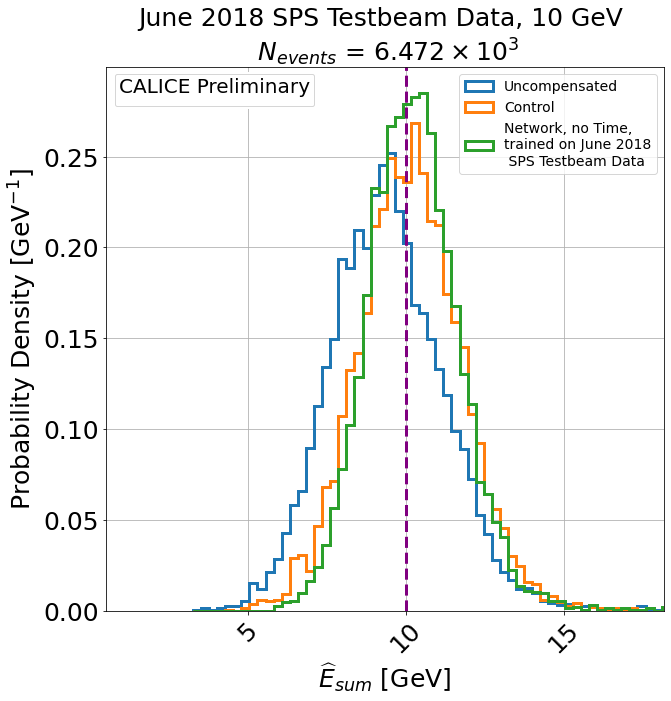}}
\hfill
\subfloat[%
 \label{fig:40GeV_Data_NoTC}
]{\includegraphics[width=0.44\linewidth]{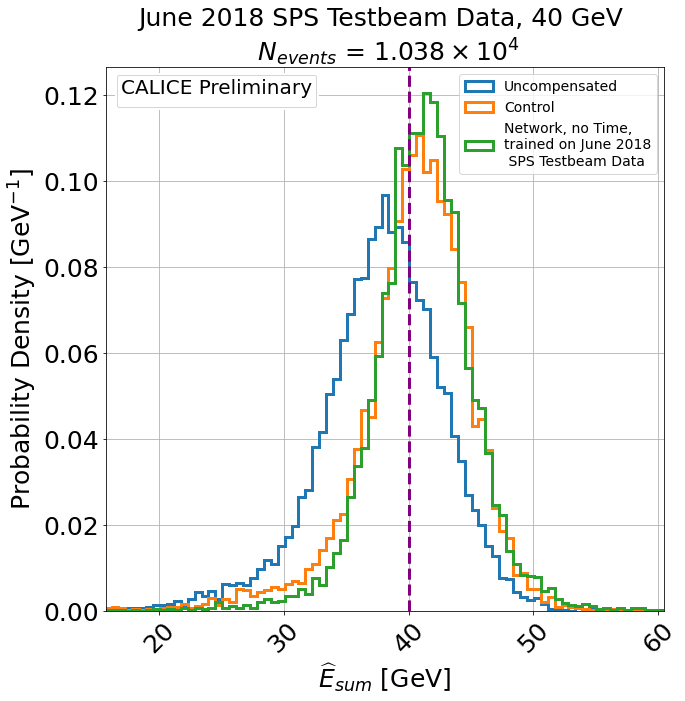}}
\vfill
\subfloat[%
 \label{fig:80GeV_Data_NoTC}
]{\includegraphics[width=0.44\linewidth]{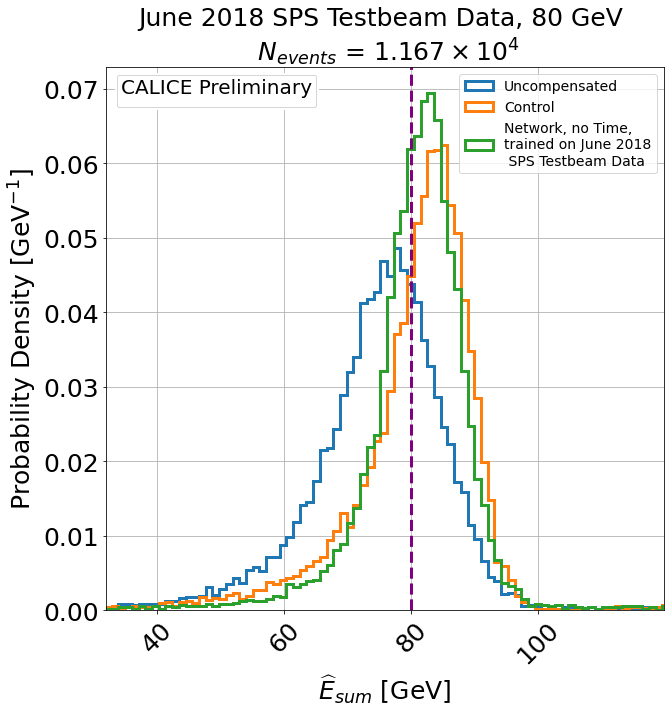}}
\hfill
\subfloat[%
 \label{fig:120GeV_Data_NoTC}
]{\includegraphics[width=0.44\linewidth]{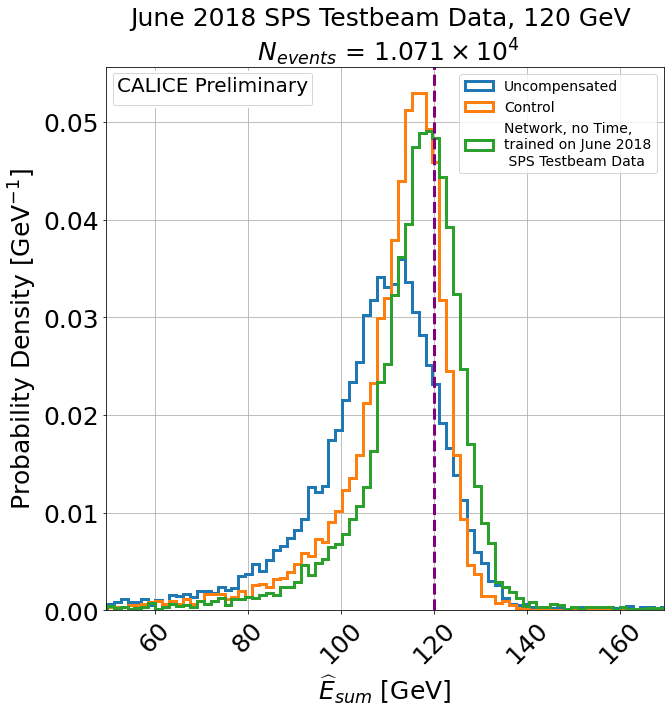}}

\caption{Example normalised histograms showing the calorimeter response before and after compensation applied to the 2018 June Testbeam test dataset of Table \ref{tab:SCNet_EventTable}, without the TCMT cut applied. Else, as in Fig.~\ref{fig:Spectra_sim} and the same selected energy samples as in Fig.~\ref{fig:Spectra_data}.}

\label{fig:Spectra_Data_NoTC}
\end{figure*}

The normalised energy response distributions for the simulation and 2018 June Testbeam test samples are shown in Figs.~\ref{fig:10GeV}-\ref{fig:120GeV} and Figs.~\ref{fig:10GeV_Data}-\ref{fig:120GeV_Data}, with the TCMT cut applied. In simulation, particle energies of \qty{10}{\giga \electronvolt}, \qty{35}{\giga \electronvolt}, \qty{80}{\giga \electronvolt} and \qty{120}{\giga \electronvolt} are shown.  For the experimental data, particle energies of \qty{10}{\giga \electronvolt}, \qty{40}{\giga \electronvolt}, \qty{80}{\giga \electronvolt} and \qty{120}{\giga \electronvolt} are shown. These samples are used for measurement of the resolution. The corresponding distributions without the TCMT cut applied are shown for the same particle energies in simulation and experimental data in Figs.~\ref{fig:10GeV_Sim_NoTC}-\ref{fig:120GeV_Sim_NoTC} and Figs.~\ref{fig:10GeV_Data_NoTC}-\ref{fig:120GeV_Data_NoTC}, respectively. The uncompensated sample and each sample after compensation are shown in each plot. The Freedman-Diaconis rule was applied to each sample to determine the bin width \cite{Binning}. The Freedman-Diaconis rule is a commonly used binning rule that approximately minimises the integral of the square difference between a histogram and a probability density function.

\paragraph{Simulation}

Fig.~\ref{fig:10GeV} shows that the neural network methods outperform the control method for the \qty{10}{\giga \electronvolt} sample, indicated by the lower spread of the response than for the control method. Furthermore, including timing information results in superior energy resolution, which is expected to play a more significant role in compensation at this energy scale due to a larger HAD fraction on average than at higher energies in the training sample, since the EM fraction increases on average with $E_{\mathrm{particle}}$ \cite{wigmans}. 

Fig.~\ref{fig:35GeV}  shows the \qty{35}{\giga \electronvolt} testing sample, which demonstrates that the neural network methods produce a more linear response than the control method and are therefore able to interpolate to samples between training energies. 

Fig.~\ref{fig:80GeV} and Fig.~\ref{fig:120GeV} show the \qty{80}{\giga \electronvolt} and \qty{120}{\giga \electronvolt} samples. The control method outperforms the neural network methods for the \qty{80}{\giga \electronvolt} sample. However, by examination of the \qty{120}{\giga \electronvolt} sample, it becomes apparent that this result is due to the control method biasing to the highest energy sample of the training dataset. This statement is justified by the artificial attenuation of the response by the control method, resulting in a highly non-linear compensated response. By contrast, the neural network methods preserve the linearity of response beyond the training range. Therefore, it is demonstrated that the neural network model can extrapolate the compensation to higher particle energies without further training.

Similar conclusions can be drawn for Figs.~\ref{fig:10GeV_Sim_NoTC}-\ref{fig:120GeV_Sim_NoTC}, indicating that the bias from the TCMT cut does not significantly influence the outcome of the experiment.

\paragraph{2018 Testbeam Data}

As in Fig.~\ref{fig:Spectra_sim}, the neural network method produces superior resolution than the control in Fig.~\ref{fig:10GeV_Data}-Fig.~\ref{fig:80GeV_Data} except for the \qty{120}{\giga \electronvolt} sample shown in Fig.~\ref{fig:120GeV_Data}. This observation can be attributed to the same energy biasing observed for the \qty{80}{\giga \electronvolt} sample in simulation in Fig.~\ref{fig:80GeV}, as these are both the maximum energy bins of the training dataset in both cases. 

Again, similar conclusions can be drawn for Figs.~\ref{fig:10GeV_Data_NoTC}-\ref{fig:120GeV_Data_NoTC}, once again indicating that the bias from the TCMT cut does not significantly influence the outcome of the experiment.

\clearpage
\subsection{Resolution and Linearity of Response}\label{sec:ResLin}

The energy response distributions for each particle energy in the testing dataset, with the TCMT cut applied, were fitted with a normal distribution in the range of $\pm 2$ standard deviations from their mean. The location and scale parameters of the fit, $\mu$ and $\sigma$, were used to estimate $E$ and $\sigma_{E}$ of Eq.~ \ref{eq:Resolution} and used to study resolution ($\sigma/\mu$ vs. $E_{\mathrm{particle}}$) and linearity of response ($\mu/E_{\mathrm{particle}}$ vs. $E_{\mathrm{particle}}$).

\paragraph{Simulation}

\begin{figure}
\centering
\includegraphics[width=0.6\linewidth]{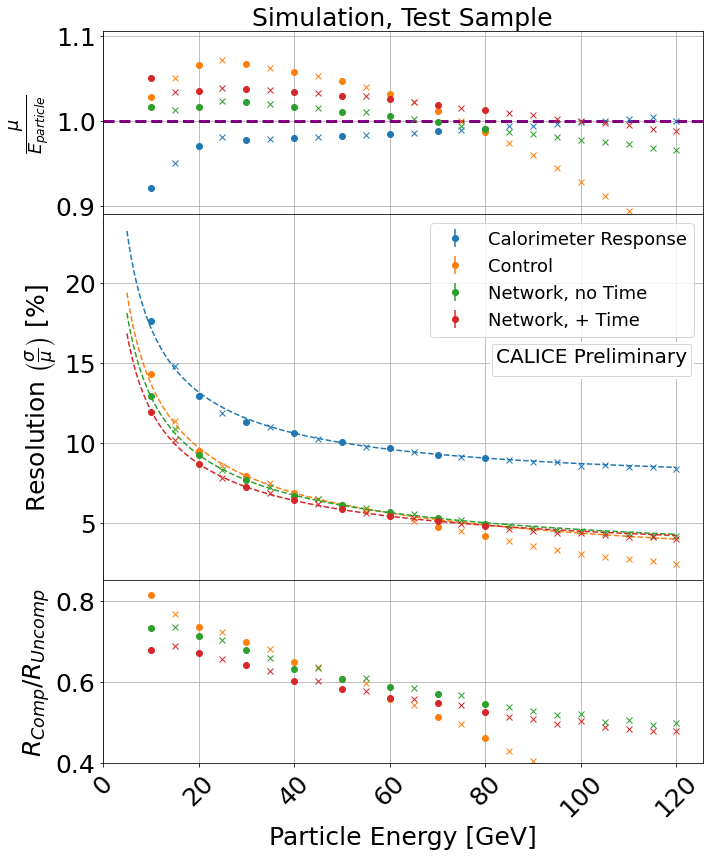}
\caption{AHCAL linearity of response and resolution using all methods under test applied to the test dataset of simulation. Blue indicates intrinsic calorimeter response, while orange, green and red indicate the control, network without and network with time compensation methods. Circles indicate energies used for both training and testing, and cross markers indicate energies used for testing only. The top subplot shows the ratio of fitted $\mu$ to $E_{\mathrm{particle}}$, where the dashed purple line indicates $\mu=E_{\mathrm{particle}}$ The middle subplot shows the fitted $\sigma/\mu$, where the dashed lines indicate fits of Eq.~\ref{eq:Resolution}. The bottom subplot indicates the ratio of the resolution of each compensation method to the intrinsic response.}
\label{fig:ResLinearity_sim}

\end{figure}

\begin{table}

    \caption{Table of fitted parameters of Eq.~\ref{eq:Resolution} to the training range of energies of simulation shown as dashed lines in Fig.~\ref{fig:ResLinearity_sim}, except for the control method, which was fitted up to $\qty{60}{\giga \electronvolt}$ due to the effect of energy biasing. }
    \label{tab:Res_sim}
    \vspace{0.2in}
    \centering
    \small{
    \begin{tabular}{l|ccc}
         \hline
         & a [\unit{\percent}] & b [\unit{\percent}] & $\chi^{2}/NDF$ \\
         \hline
         Uncompensated & $49.5\pm 0.4$ &  $7.1 \pm 0.1$ & $4.6$ \\
         Control & $43.4 \pm 0.1$ & $0.0 \pm 2.9$ & $14.3$ \\
         Network, No Time &  $40.2 \pm 0.2$ & $2.2 \pm 0.1$ &  $0.9$ \\
         Network, + Time &  $37.3 \pm 0.2$ & $2.4 \pm 0.1$ & $1.4$ \\
         \hline
    \end{tabular}
    }

\end{table}

Fig.~\ref{fig:ResLinearity_sim} and Table \ref{tab:Res_sim} show the measured resolution, fitted with Eq.~\ref{eq:Resolution}, and the corresponding fit values, respectively, for the simulation.

The top subplot of Fig.~\ref{fig:ResLinearity_sim} indicates the neural network methods offer improved linearity of response compared to the control, which overestimates the hadron shower energy by up to $\qty{5}{\percent}$ compared to $2$-$\qty{3}{\percent}$ for the network methods for most of the training range of particle energies. Moreover, the network and control methods are demonstrated to interpolate within the training range. However, the control method fails to reconstruct the particle energy entirely for particle energies greater than $\qty{80}{\giga \electronvolt}$. 


The middle and bottom subfigure of Fig.~\ref{fig:ResLinearity_sim} demonstrates that for values of $E_{\mathrm{particle}}$ up to around $\qty{60}{\giga \electronvolt}$, the neural network methods produce superior compensation, indicated by the smaller value of the compensated response to the intrinsic response. Beyond this range, the resolution produced by the control method diverges from the model of Eq.~\ref{eq:Resolution}. For this reason, the fit to this method was only performed for $E_{\mathrm{particle}}$ in the range $10$-$\qty{60}{\giga \electronvolt}$. By contrast, the uncompensated and network methods show good agreement with the expectations of Eq.~\ref{eq:Resolution} and were fitted over the entire range. Table \ref{tab:Res_sim} shows the fitted parameters indicated by the dashed coloured lines of the middle subplot of Fig.~\ref{fig:ResLinearity_sim}. The uncompensated stochastic resolution for simulated $\pi^{-}$ hadron showers in AHCAL is in agreement within $1$-$\qty{2}{\percent}$ with the $a=51.7 \pm \qty{0.97}{\percent}$ obtained in the study of \cite{Olin}. The neural network solutions improve the calorimeter's stochastic resolution, $a$, by comparison by around $\qty{3}{\percent}$ without timing information and a further $\qty{3}{\percent}$ with timing information, compared to the control method. This result demonstrates the improvement in SC performance that can be obtained from including spatiotemporal energy density information. This result agrees with a similar study on the additional benefit of using timing information for software compensation with AHCAL, which observed a 3-\qty{4}{\percent} improvement in energy resolution using timing information than energy density information only.

\paragraph{June 2018 Testbeam Data}

\begin{figure}
\centering
\includegraphics[width=0.6\linewidth]{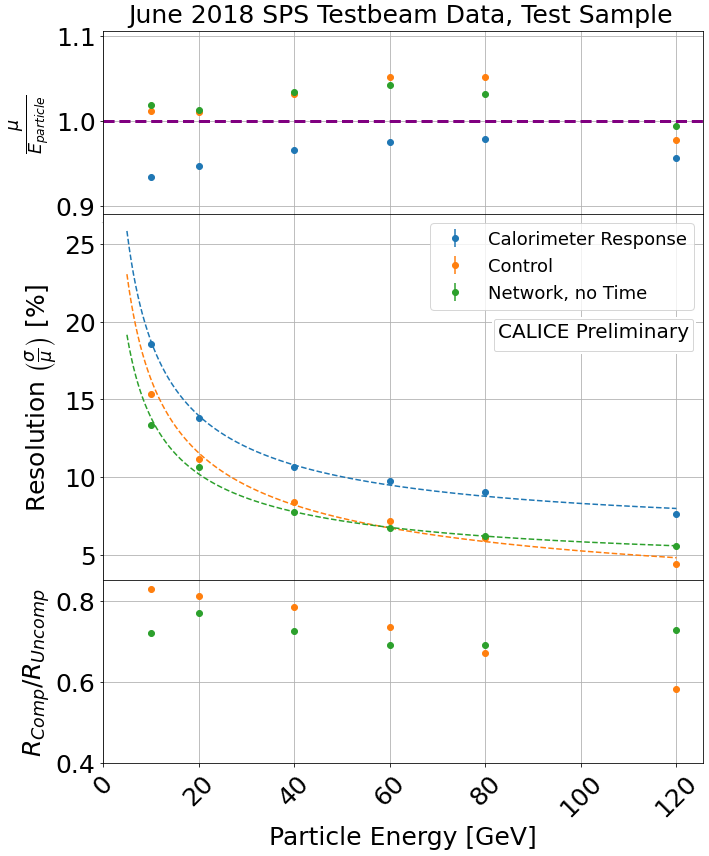}
\caption{AHCAL linearity of response and resolution using all methods under test applied to the test dataset of 2018 CALICE Testbeam Data. Else, as in Fig.~\ref{fig:ResLinearity_sim}.}
\label{fig:ResLinearity_data}

\end{figure}

Fig.~\ref{fig:ResLinearity_data} and Table \ref{tab:Res_data} show the measured resolution, fitted with Eq.~\ref{eq:Resolution}, and the corresponding fit values, respectively, for the data.

As for simulation, the uncompensated stochastic resolution for data $\pi^{-}$ hadron showers in AHCAL is in agreement within 1-\qty{2}{\percent} with the stochastic resolution term $a = 57.70 \pm \qty{1.06}{\percent}$ obtained in \cite{Olin}. Furthermore, the machine learning methods outperform the control method in resolution, resulting in an improvement of the intrinsic stochastic resolution term of \qty{9.3}{\percent} and \qty{12.2}{\percent}, outperforming the control method in both cases. The neural networks also reduce the constant resolution term by \qty{2}{\percent}, indicating the neural networks perform some detector calibration and SC. A slightly superior linearity of response overall is observed compared to the control method and less than for simulation.

\begin{table*}
    \caption{Table of fitted parameters of Eq.~\ref{eq:Resolution} to the training range of energies of 2018 CALICE Testbeam data shown as dashed lines in Fig.~\ref{fig:ResLinearity_sim}. The whole range of available energies was used to fit. }
    \label{tab:Res_data}
    \vspace{0.2in}
    \centering
    \small{
    \begin{tabular}{l|ccc}
         \toprule
         & a [\unit{\percent}] & b [\unit{\percent}] & $\chi^{2}/NDF$ \\
         \midrule
         Calorimeter Response & $56.1\pm 0.7$ &  $6.1 \pm 0.1$ & $ 10.1$ \\
         Control & $51.5 \pm 0.42$ & $1.0 \pm 0.3$ & $38.9$ \\
         Network, No Time &  $41.9\pm 0.5$ & $4.0 \pm 0.1$ &  $6.5$\\
         \bottomrule
    \end{tabular}
    }

\end{table*}

\subsection{Correlations with Spatial and Temporal Information}

The spatial and energy-temporal correlations of the hadron shower weighting are analysed to study and compare the neutral network SC methods to the control SC method.

\begin{figure*}
\vspace{-0.8in}
\subfloat[%
\label{fig:Spatial_Control}
]{\includegraphics[width=0.4\linewidth]{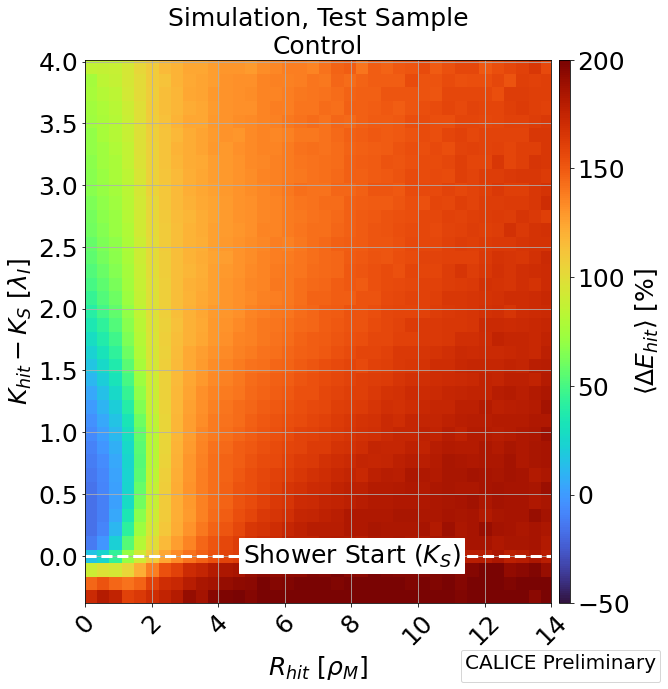}}
\hfill
\subfloat[%
\label{fig:Energy_Control}
]{\includegraphics[width=0.4\linewidth]{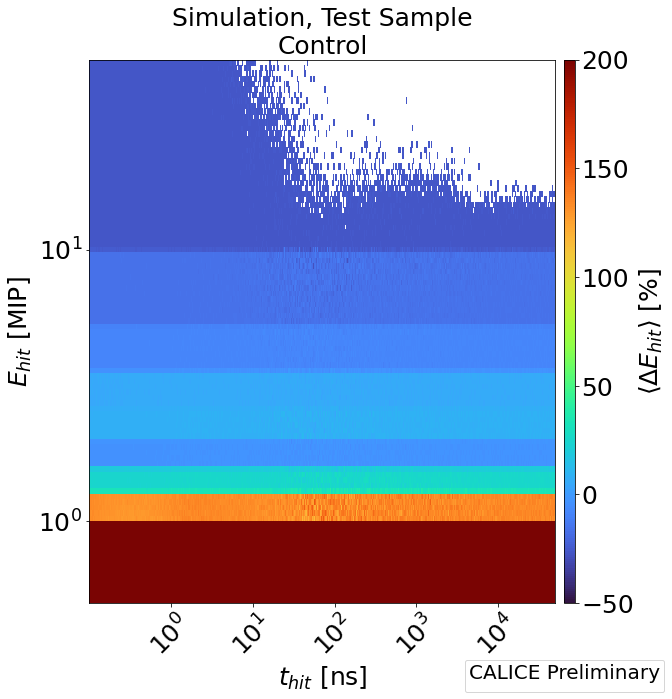}}

\subfloat[%
\label{fig:Spatial_NetNoTime}
]{\includegraphics[width=0.4\linewidth]{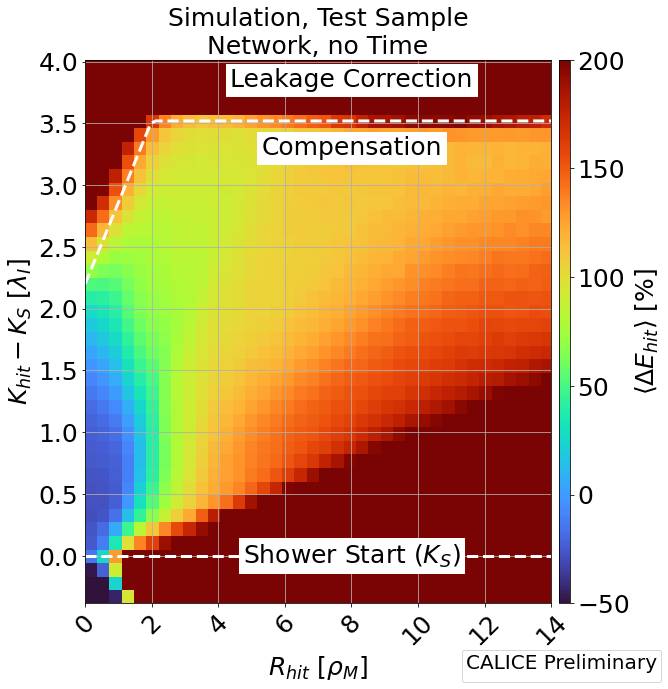}}
\hfill
\subfloat[%
\label{fig:Energy_NetNoTime}
]{\includegraphics[width=0.4\linewidth]{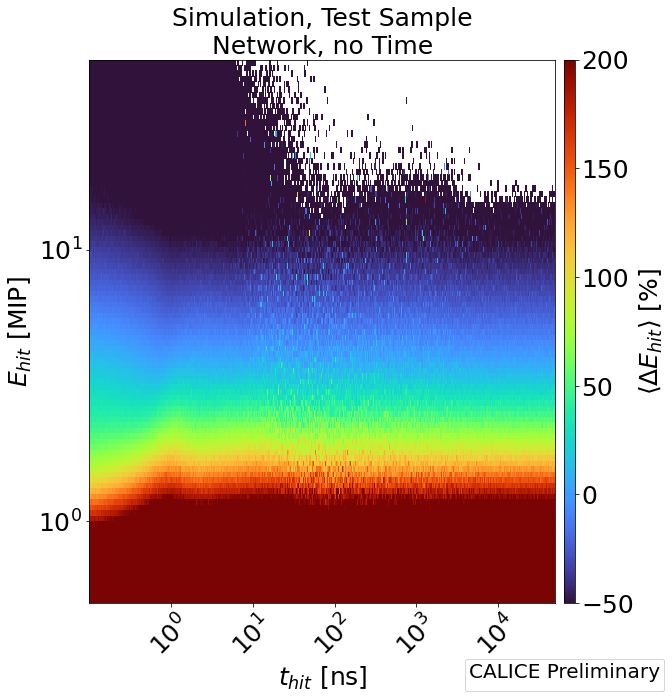}}

\subfloat[%
\label{fig:Spatial_NetwTime}
]{\includegraphics[width=0.4\linewidth]{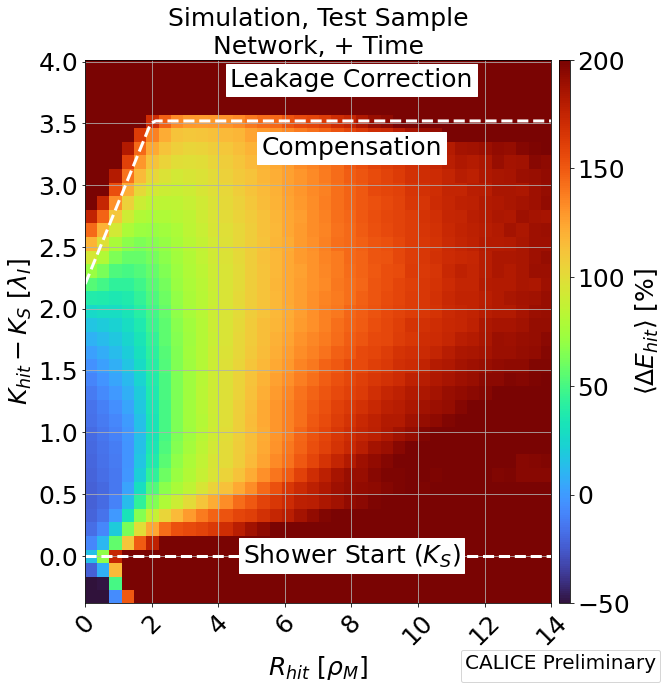}}
\hfill
\subfloat[%
\label{fig:Energy_NetwTime}
]{\includegraphics[width=0.4\linewidth]{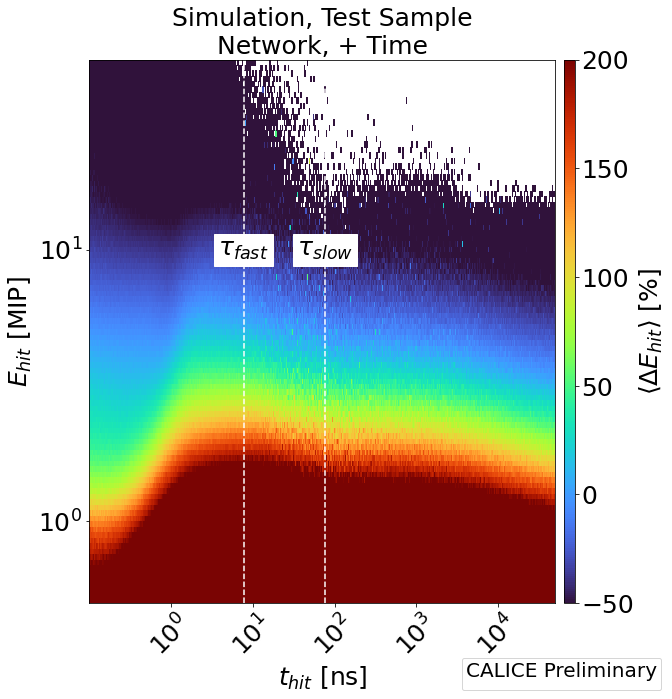}}

\caption{Average percentage change in active cell energy ($E_{\mathrm{hit}}$) as a result of compensation as a function of $R_{\mathrm{hit}}$ and $K_{\mathrm{hit}} - K_{S}$ (left column, presented in units of Moliere radius from the lateral center-of-gravity, $\rho_{M} = \qty{24.9}{\milli \meter}$, and nuclear interaction length from the shower start, $\lambda_{I} = \qty{237.1}{\milli \meter}$, respectively), and $E_{\mathrm{hit}}$ and $t_{\mathrm{hit}}$ (right column, presented in units of \unit{\text{MIP}} and \unit{\nano \second}, respectively) for simulation. Each row indicates the control and network methods without and with timing information in that order. The colour axis indicates the percentage change, where blue regions indicate where the energy has been attenuated, and green through red shows where the energy has been enhanced.  White space indicates no data available. Regions of interest are labelled accordingly for reference. The values of $\tau_{\mathrm{slow}}$ and $\tau_{\mathrm{fast}}$ were taken from \cite{time}.}
\label{fig:Correlations}
\end{figure*}

\paragraph{Simulation}

The results for the spatial correlations ($R_{\mathrm{hit}}, K_{\mathrm{hit}} - K_{S}$) and energy-temporal correlations  ($t_{\mathrm{hit}}, E_{\mathrm{hit}}$) are shown in the left and right columns of Fig.~\ref{fig:Correlations}, respectively to the test sample. The colour axes indicate the percentage change of the energy due to the SC algorithm as a function of these variables. In this example, the tail-catcher cut was not applied. 

 Fig.~\ref{fig:Spatial_Control} demonstrates that the control method shows only a weak dependence on lateral and longitudinal development of the shower, with attenuation occurring only within $R_{\mathrm{hit}} \lesssim  \qty{1}{\text{\ensuremath{\rho_{M}}}}$ (the EM fraction) and enhancement beyond, with minor variation, as expected. By contrast, the neural network methods attenuate and enhance the active cell energy with much stronger spatial dependence, indicated by the broadening of the weighting profile with longitudinal shower development. Two additional effects are observed for the network methods, shown in Fig.~\ref{fig:Spatial_NetNoTime} and Fig.~\ref{fig:Spatial_NetwTime}: a tendency to enhance $E_{\mathrm{hit}}$ in the region above the white dashed line, and to attenuate $E_{\mathrm{hit}}$ where $R_{\mathrm{hit}} \lesssim  1 \rho_{M}$ (close to the lateral shower core) and $K_{S}<0$ (before the shower start). These effects are not present in Fig.~\ref{fig:Spatial_Control} and must therefore be a consequence of including spatial information in the models. These results suggest the network models have learned leakage correction and to remove the energy deposited by minimum ionisation of the $\pi^{-}$ particle before showering. This result demonstrates an improved capacity of the proposed model to learn the physical properties of the hadron shower and detector compared to the control method.

Fig.~\ref{fig:Energy_Control},  Fig.~\ref{fig:Energy_NetNoTime} and Fig.~\ref{fig:Energy_NetwTime} demonstrate that all methods are observed to attenuate active cell energies above $\qty{5}{\text{MIP}}$ and enhance below that threshold, which is expected of all SC algorithms. The binned structure of the weighting of the control method is visible in Fig.~\ref{fig:Energy_Control}. By contrast, the neural network methods in Fig.~\ref{fig:Energy_NetNoTime} and  Fig.~\ref{fig:Energy_NetwTime} indicate a continuous weighting function has been learned. Furthermore, Fig.~\ref {fig:Energy_NetwTime} indicates that the model with timing information enhances the threshold for energy deposited in the order of several \unit{\nano \second} to several tens of \unit{\nano \second}. A reduction in the threshold is observed after around \qty{100}{\nano \second}. These observations are consistent with the timescales of the two main neutron energy-depositing processes discussed in Section \ref{sec:CALICE_AHCAL}. Comparison of Fig.~\ref{fig:Energy_NetwTime} and Figs.~\ref{fig:Energy_Control}-~\ref{fig:Energy_NetNoTime} indicate that this effect must be due to the inclusion of timing information since no such effect is observed in the control or method without timing information.

\paragraph{CALICE 2018 Testbeam Data}

\begin{figure*}
\vspace{-0.8in}
\subfloat[%
\label{fig:Spatial_Control_Data}
]{\includegraphics[width=0.4\linewidth]{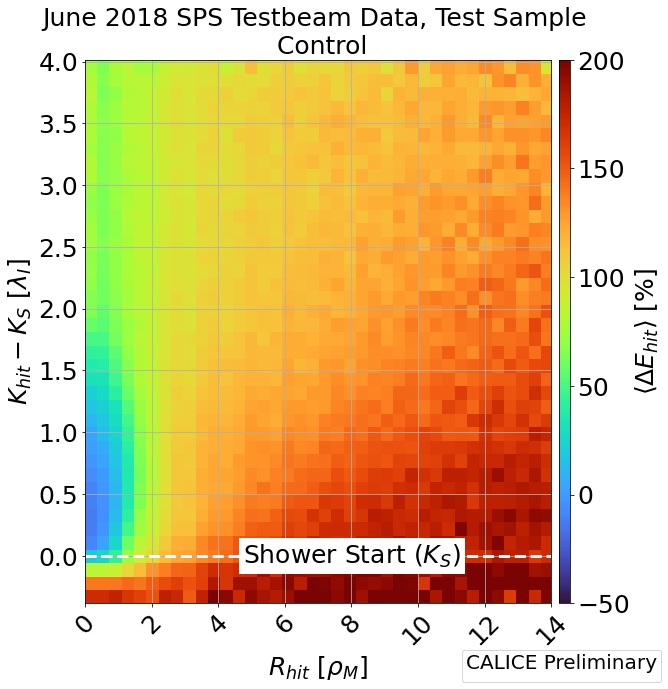}}
\hfill
\subfloat[%
\label{fig:Energy_Control_Data}
]{\includegraphics[width=0.4\linewidth]{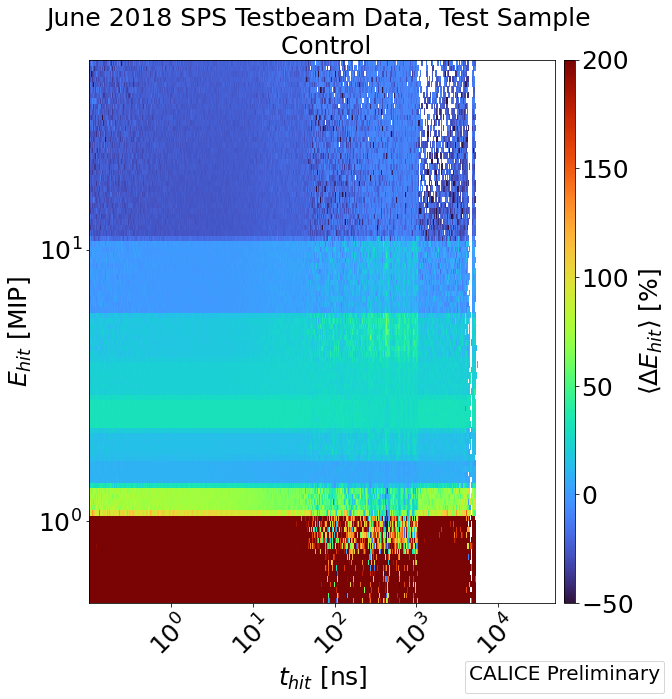}}

\subfloat[%
\label{fig:Spatial_NetNoTime_Data}
]{\includegraphics[width=0.4\linewidth]{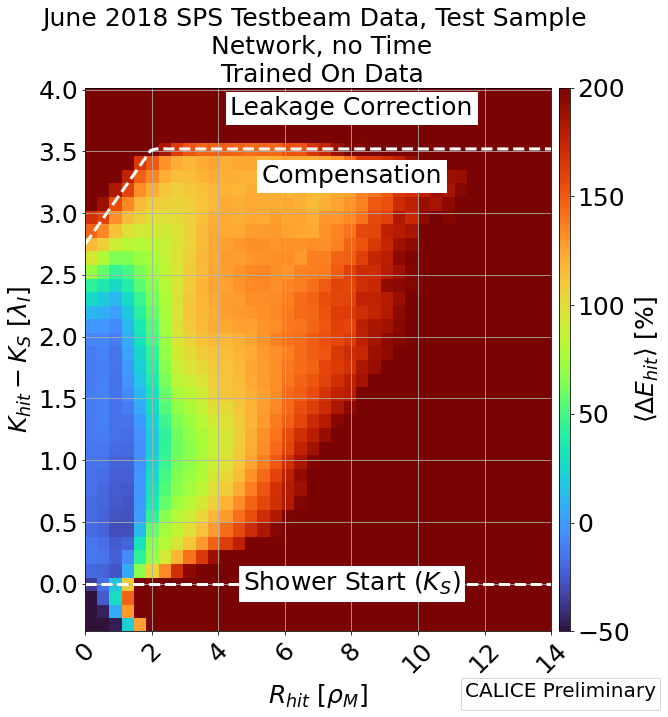}}
\hfill
\subfloat[%
\label{fig:Energy_NetNoTime_Data}
]{\includegraphics[width=0.4\linewidth]{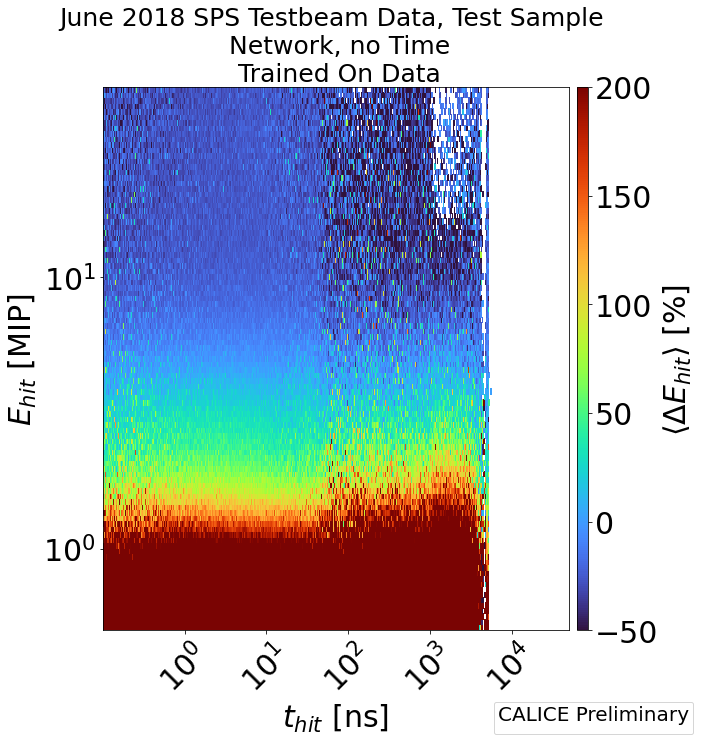}}


\caption{Average percentage change in active cell energy ($E_{\mathrm{hit}}$) as a result of compensation as a function of $R_{\mathrm{hit}}$ and $K_{\mathrm{hit}} - K_{S}$ as in Fig.~\ref{fig:Correlations} for CALICE 2018 Testbeam Data. Else, as in Fig.~\ref{fig:Correlations}.}
\label{fig:Correlations_data}
\end{figure*}

The spatial and energy-temporal correlations in Fig.~\ref{fig:Correlations_data} show the same information for data as in Figs.~\ref{fig:Spatial_Control}-\ref{fig:Energy_NetNoTime}. The conclusions to this figure are the same as in Fig.~\ref{fig:Correlations} for the control method and the neural network model trained only with spatial information.

\clearpage

\section{Conclusion}

A neural network method for performing software compensation was devised, trained, and tested on simulation and 2018 June Testbeam data for the AHCAL calorimeter. The model used a local energy density estimate to overcome biasing effects on particle energies.

The neural network model was trained with and without timing information with \qty{100}{\pico \second} timing resolution and is compared to a control method after accounting for the effect of leakage compensation learned by the networks. The neural networks yielded superior overall compensation and linearity of response to the control method when trained on simulation, resulting in calorimeter resolutions of $\qty{40.2}{\percent}/\sqrt{E_{\mathrm{particle}}} \oplus \qty{2.2}{\percent}$ and  $\qty{37.2}{\percent}/\sqrt{E_{\mathrm{particle}}} \oplus \qty{2.4}{\percent}$. This corresponded to an absolute improvement of stochastic resolution, $a$, by \qty{9.3}{\percent} and \qty{12.2}{\percent}, or a relative improvement of \qty{19}{\percent} and \qty{25}{\percent} respectively compared to the uncompensated $a$. The absolute value of the constant resolution term, $b$, was also found to reduce by around \qty{5}{\percent} or a relative improvement of \qty{70}{\percent} compared to the uncompensated $b$ in both models. This result indicates that the model was capable of detector calibration.  Both methods obtained a linearity of response within around $2$-$\qty{3}{\percent}$ of the particle energy. This result should be interpreted with the caveat of the bias caused by the TCMT cut. Nonetheless, improved performance was observed compared to the control using the neural network method without the TCMT cut.

The network without timing information trained on CALICE 2018 SPS testbeam data achieved a comparable resolution of  $\qty{41.9}{\percent}/\sqrt{E_{\mathrm{particle}}} \oplus \qty{4.0}{\percent}$. This corresponded to an absolute improvement in $a$ by \qty{14.2}{\percent} or a relative improvement of \qty{25}{\percent}. Additionally, this corresponded to an absolute improvement in $b$ by around \qty{2}{\percent} or a relative improvement of around \qty{35}{\percent}. This result indicates that the model can be trained with limited experimental data to a similar level as simulation. Additionally, the control method was observed to bias to the training range of particle energies. In contrast, the neural network method was demonstrated to both interpolate and extrapolate compensation to energies not used for training.

The networks' learned method of applying SC agreed with expectations: the attenuation of high-energy (EM) deposits and the enhancement of low-energy deposits. However, the network method was found to apply SC differently depending on the stage of the shower development, both in space and in time, the former of which included an effect consistent with longitudinal leakage correction and the latter of which was found to agree with expectations of a bi-exponential time distribution for energy deposits in a steel-scintillator calorimeter expected from Ref.~\cite{time}. Similar behaviour was observed in the independent neural network applied to data.

In summary, this study indicates that superior resolution can be obtained in highly granular calorimeters using spatiotemporal event information and neural networks and that careful model design can overcome the limitations of previous data-driven compensation techniques by reducing energy biasing. The validation of the method as part of a full Particle Flow analysis represents a promising future study. 

\clearpage 
\appendix
\setcounter{secnumdepth}{0}
\section{Appendix}

\begin{table*}[htb!]

    \caption[Bin Ranges and Weights obtained for the SC Control Method]{Bin ranges and weights obtained for the control method described in Section \ref{sec:ControlMethod}. Table \ref{tab:SCNet_LSCWeightsSim} shows the values obtained for simulation. Table \ref{tab:SCNet_LSCWeightsData} shows the values obtained for CALICE June 2018 SPS Testbeam data.}
    \label{tab:LSCWeights}
    \vspace{0.2in}

    \centering

    \subfloat[\label{tab:SCNet_LSCWeightsSim}]{
    \scriptsize{
    \begin{tabular}{llll}
    \toprule
     &  \multicolumn{3}{l}{Simulation}  \\
    Bin Range [\unit{\text{MIP}}] & $\alpha_{\mathrm{b}}$ & $\beta_{\mathrm{b}}$ & $\gamma_{\mathrm{b}}$ \\
    \midrule
     0.500 - 0.735 &  -1.120 &  13.000 &  -5.499 \\
     0.735 - 1.002 &  -0.372 &   5.134 &  -4.727 \\
     1.002 - 1.272 &  -0.350 &   2.523 &  -3.054 \\
     1.272 - 1.585 &  -0.460 &   1.999 &  -1.587 \\
     1.585 - 2.013 &  -0.290 &   1.634 &  -1.012 \\
     2.013 - 2.631 &   0.022 &   1.275 &  -0.949 \\
     2.631 - 3.584 &   0.218 &   1.138 &  -0.632 \\
     3.584 - 5.328 &   0.387 &   0.658 &  -0.457 \\
     5.328 - 9.881 &   0.579 &   0.259 &  -0.281 \\
     9.881 - $\infty$ &   0.788 &  -0.057 &   0.043 \\
    \bottomrule
    \end{tabular}
    }
    }
    \hfill
    \subfloat[\label{tab:SCNet_LSCWeightsData}]{
    \scriptsize{
    \begin{tabular}{llll}
    \toprule
    & \multicolumn{3}{l}{June 2018 SPS Testbeam Data} \\
    Bin Range [\unit{\text{MIP}}] & $\alpha_{\mathrm{b}}$ & $\beta_{\mathrm{b}}$ & $\gamma_{\mathrm{b}}$ \\
    \midrule
    0.500 - 0.770 &  -9.252 &  17.803 &  -16.047 \\
    0.770 - 1.059 &  -8.529 &  14.738 &  -11.256 \\
    1.059 - 1.351 &  -0.046 &   1.902 &   -1.797 \\
    1.351 - 1.698 &   1.283 &  -0.379 &    0.057 \\
    1.698 - 2.179 &   2.104 &  -1.325 &    0.720 \\
    2.179 - 2.875 &   1.025 &   0.384 &   -0.235 \\
    2.875 - 3.957 &   1.271 &   0.001 &    0.104 \\
    3.957 - 5.847 &   1.325 &   0.013 &    0.304 \\
    5.847 - 10.930 &   0.814 &   0.426 &    0.021 \\
    10.930 - $\infty$ &   0.148 &   1.032 &   -0.243 \\
    \bottomrule
    \end{tabular}
   
    }
    }

\end{table*}

\begin{table*}[htb!]

\centering
\caption[Tables of $\mu$ and $\sigma$ and reduced $\chi^{2}$ From Gaussian Fits Performed on SC models For Resolution Measurements of Simulation]{Table of $\mu$ and $\sigma$ from the Gaussian fits performed on the SC models trained on simulation. Tables \ref{tab:SCNet_Mu_Stat_Sim} and \ref{tab:SCNet_Sigma_Stat_Sim}  show the $\mu$, $\sigma$ and their errors as a function of particle energy for each studied method applied to the testing dataset. CR, CTRL, NN,-Time and NN,+Time are abbreviations of: 'intrinsic calorimeter response', 'control method', 'neural network, without time' and with 'neural network, with time', respectively.}
\label{tab:SCNet_ResLin_Stats_Sim}
\vspace{0.2in}
\subfloat[\label{tab:SCNet_Mu_Stat_Sim}]{
    \centering
    \tiny{
   \begin{tabular}{lrrrrrrrr}
    \toprule
    {} & \multicolumn{8}{l}{Simulation} \\
    {} & \multicolumn{4}{l}{$\mu$} & \multicolumn{4}{l}{$\mathrm{d}\mu$} \\
    & CR & CTRL & NN,-Time & NN,+Time & CR & CNTRL & NN,-Time & NN,+ Time \\
    $E_{\mathrm{particle}}$ [\unit{\giga \electronvolt}] &                   &                       &                               &                              &                   &                       &                               &                              \\
    \midrule
    10             &             9.214 &                10.281 &                        10.157 &                       10.502 &             0.013 &                 0.012 &                         0.010 &                        0.009 \\
    15             &            14.257 &                15.754 &                        15.186 &                       15.512 &             0.016 &                 0.014 &                         0.012 &                        0.012 \\
    20             &            19.401 &                21.307 &                        20.326 &                       20.692 &             0.019 &                 0.015 &                         0.013 &                        0.013 \\
    25             &            24.516 &                26.796 &                        25.569 &                       25.954 &             0.021 &                 0.016 &                         0.015 &                        0.014 \\
    30             &            29.330 &                31.990 &                        30.661 &                       31.136 &             0.024 &                 0.018 &                         0.017 &                        0.016 \\
    35             &            34.230 &                37.160 &                        35.688 &                       36.247 &             0.027 &                 0.020 &                         0.018 &                        0.018 \\
    40             &            39.199 &                42.311 &                        40.656 &                       41.368 &             0.030 &                 0.021 &                         0.019 &                        0.019 \\
    45             &            44.148 &                47.374 &                        45.651 &                       46.468 &             0.034 &                 0.023 &                         0.022 &                        0.021 \\
    50             &            49.119 &                52.316 &                        50.526 &                       51.466 &             0.034 &                 0.022 &                         0.021 &                        0.021 \\
    55             &            54.066 &                57.157 &                        55.555 &                       56.583 &             0.043 &                 0.028 &                         0.027 &                        0.026 \\
    60             &            59.033 &                61.864 &                        60.352 &                       61.535 &             0.045 &                 0.027 &                         0.026 &                        0.026 \\
    65             &            64.099 &                66.412 &                        65.148 &                       66.438 &             0.045 &                 0.026 &                         0.026 &                        0.026 \\
    70             &            69.145 &                70.838 &                        69.905 &                       71.315 &             0.057 &                 0.031 &                         0.032 &                        0.032 \\
    75             &            74.178 &                74.941 &                        74.578 &                       76.158 &             0.066 &                 0.035 &                         0.037 &                        0.037 \\
    80             &            79.235 &                78.972 &                        79.270 &                       80.992 &             0.061 &                 0.029 &                         0.033 &                        0.032 \\
    85             &            84.506 &                82.781 &                        83.912 &                       85.779 &             0.064 &                 0.029 &                         0.034 &                        0.033 \\
    90             &            89.475 &                86.410 &                        88.615 &                       90.592 &             0.063 &                 0.027 &                         0.033 &                        0.032 \\
    95             &            94.663 &                89.725 &                        93.157 &                       95.238 &             0.078 &                 0.031 &                         0.039 &                        0.039 \\
    100            &            99.804 &                92.822 &                        97.723 &                       99.969 &             0.088 &                 0.034 &                         0.044 &                        0.044 \\
    105            &           104.983 &                95.722 &                       102.337 &                      104.685 &             0.095 &                 0.032 &                         0.047 &                        0.047 \\
    110            &           110.224 &                98.338 &                       106.931 &                      109.416 &             0.103 &                 0.031 &                         0.050 &                        0.049 \\
    115            &           115.567 &               100.585 &                       111.316 &                      113.949 &             0.106 &                 0.028 &                         0.050 &                        0.049 \\
    120            &           120.000 &               102.513 &                       115.874 &                      118.618 &             0.010 &                 0.028 &                         0.053 &                        0.052 \\
\bottomrule
\end{tabular}
}

}

\subfloat[\label{tab:SCNet_Sigma_Stat_Sim}]{

\tiny{
\begin{tabular}{lrrrrrrrr}
    \toprule
    {} & \multicolumn{8}{l}{Simulation} \\
    {} & \multicolumn{4}{l}{$\sigma$} & \multicolumn{4}{l}{$\mathrm{d}\sigma$} \\
    & CR & CTRL & NN,-Time & NN,+Time & CR & CNTRL & NN,-Time & NN,+ Time \\
    $E_{\mathrm{particle}}$ [\unit{\giga \electronvolt}] &                   &                       &                               &                              &                   &                       &                               &                              \\
\midrule
10             &             1.625 &                 1.474 &                         1.311 &                        1.255 &             0.012 &                 0.011 &                         0.009 &                        0.008 \\
15             &             2.112 &                 1.792 &                         1.651 &                        1.579 &             0.015 &                 0.012 &                         0.010 &                        0.010 \\
20             &             2.512 &                 2.027 &                         1.873 &                        1.795 &             0.018 &                 0.013 &                         0.011 &                        0.011 \\
25             &             2.905 &                 2.294 &                         2.128 &                        2.016 &             0.019 &                 0.013 &                         0.013 &                        0.012 \\
30             &             3.320 &                 2.527 &                         2.351 &                        2.258 &             0.021 &                 0.015 &                         0.014 &                        0.013 \\
35             &             3.754 &                 2.779 &                         2.576 &                        2.487 &             0.023 &                 0.016 &                         0.015 &                        0.015 \\
40             &             4.151 &                 2.905 &                         2.721 &                        2.642 &             0.025 &                 0.017 &                         0.016 &                        0.015 \\
45             &             4.518 &                 3.081 &                         2.961 &                        2.858 &             0.027 &                 0.018 &                         0.018 &                        0.018 \\
50             &             4.946 &                 3.192 &                         3.092 &                        3.023 &             0.027 &                 0.018 &                         0.017 &                        0.017 \\
55             &             5.253 &                 3.316 &                         3.296 &                        3.182 &             0.033 &                 0.022 &                         0.023 &                        0.022 \\
60             &             5.701 &                 3.335 &                         3.416 &                        3.328 &             0.034 &                 0.021 &                         0.022 &                        0.022 \\
65             &             6.059 &                 3.406 &                         3.598 &                        3.510 &             0.034 &                 0.021 &                         0.022 &                        0.022 \\
70             &             6.406 &                 3.367 &                         3.695 &                        3.628 &             0.044 &                 0.023 &                         0.027 &                        0.027 \\
75             &             6.736 &                 3.376 &                         3.840 &                        3.751 &             0.051 &                 0.027 &                         0.032 &                        0.032 \\
80             &             7.174 &                 3.297 &                         3.917 &                        3.862 &             0.046 &                 0.022 &                         0.027 &                        0.027 \\
85             &             7.540 &                 3.167 &                         4.021 &                        3.925 &             0.048 &                 0.021 &                         0.029 &                        0.029 \\
90             &             7.843 &                 3.078 &                         4.111 &                        4.042 &             0.045 &                 0.020 &                         0.028 &                        0.028 \\
95             &             8.297 &                 2.943 &                         4.237 &                        4.141 &             0.059 &                 0.021 &                         0.033 &                        0.034 \\
100            &             8.553 &                 2.803 &                         4.354 &                        4.318 &             0.062 &                 0.023 &                         0.036 &                        0.037 \\
105            &             9.049 &                 2.741 &                         4.425 &                        4.422 &             0.071 &                 0.023 &                         0.039 &                        0.039 \\
110            &             9.341 &                 2.664 &                         4.588 &                        4.488 &             0.078 &                 0.024 &                         0.044 &                        0.041 \\
115            &             9.832 &                 2.603 &                         4.666 &                        4.643 &             0.081 &                 0.026 &                         0.040 &                        0.041 \\
120            &            10.000 &                 2.493 &                         4.813 &                        4.742 &             0.027 &                 0.028 &                         0.045 &                        0.044 \\
\bottomrule
\end{tabular}

}

}

\end{table*}

\begin{table*}[htb!]
\caption[Tables of $\mu$ and $\sigma$ and reduced $\chi^{2}$ From Gaussian Fits Performed on SC models For Resolution Measurements of June 2018 SPS Testbeam Data]{Table of $\mu$ and $\sigma$ from the Gaussian fits performed on the SC models trained on data. Tables \ref{tab:SCNet_Mu_Stat_Data},  \ref{tab:SCNet_Sigma_Stat_Data} $\mu$, $\sigma$ and their errors as a function of particle energy for each studied method applied to the testing dataset. Else, as in Table \ref{tab:SCNet_ResLin_Stats_Sim}. 'Sim' and 'Data' indicate models trained on simulation and 2018 Testbeam data, resepctively.}
\label{tab:SCNet_ResLin_Stats_Data}
\vspace{0.2in}
\centering
    \subfloat[\label{tab:SCNet_Mu_Stat_Data}]{
    \centering
    \tiny{
   \begin{tabular}{lrrrrrrrr}
      \toprule
    {} & \multicolumn{8}{l}{June 2018 SPS Testbeam Data} \\
    {} & \multicolumn{4}{l}{$\mu$} & \multicolumn{4}{l}{$\mathrm{d}\mu$} \\
    & CR & CTRL & NN,-Time & NN,-Time & CR & CNTRL & NN,-Time & NN,-Time \\
    $E_{\mathrm{particle}}$ [\unit{\giga \electronvolt}] &                   &                 (Data)       &                         (Data)       &        (Sim)                      &                  &                  (Data)       &                        (Data)      &         (Sim)                     \\
\midrule
10             &             9.334 &                10.113 &                        10.190 &                             9.949 &             0.025 &                 0.022 &                         0.018 &                             0.019 \\
20             &            18.947 &                20.213 &                        20.244 &                            19.048 &             0.031 &                 0.026 &                         0.025 &                             0.024 \\
40             &            38.623 &                41.278 &                        41.355 &                            38.301 &             0.047 &                 0.038 &                         0.036 &                             0.033 \\
60             &            58.501 &                63.090 &                        62.503 &                            57.331 &             0.060 &                 0.047 &                         0.043 &                             0.038 \\
80             &            78.246 &                84.081 &                        82.514 &                            75.260 &             0.081 &                 0.062 &                         0.058 &                             0.049 \\
120            &           114.690 &               117.347 &                       119.229 &                           107.497 &             0.121 &                 0.080 &                         0.092 &                             0.077 \\
    \bottomrule
\end{tabular}
}
}

\subfloat[\label{tab:SCNet_Sigma_Stat_Data}]{

\tiny{
\begin{tabular}{lrrrrrrrr}
    \toprule
    {} & \multicolumn{8}{l}{June 2018 SPS Testbeam Data} \\
    {} & \multicolumn{4}{l}{$\sigma$} & \multicolumn{4}{l}{$\mathrm{d}\sigma$} \\
    & CR & CTRL & NN,-Time & NN,-Time & CR & CNTRL & NN,-Time & NN,-Time \\
    $E_{\mathrm{particle}}$ [\unit{\giga \electronvolt}] &                   &                 (Data)       &                         (Data)       &        (Sim)                      &                  &                  (Data)       &                        (Data)      &         (Sim)                     \\
\midrule
10             &             1.733 &                 1.554 &                         1.363 &                             1.407 &             0.023 &                 0.019 &                         0.016 &                             0.016 \\
20             &             2.614 &                 2.260 &                         2.151 &                             2.094 &             0.029 &                 0.023 &                         0.022 &                             0.021 \\
40             &             4.124 &                 3.456 &                         3.198 &                             3.012 &             0.042 &                 0.033 &                         0.031 &                             0.029 \\
60             &             5.702 &                 4.526 &                         4.202 &                             3.750 &             0.050 &                 0.039 &                         0.036 &                             0.031 \\
80             &             7.058 &                 5.093 &                         5.144 &                             4.381 &             0.065 &                 0.046 &                         0.048 &                             0.040 \\
120            &             8.734 &                 5.203 &                         6.611 &                             5.532 &             0.096 &                 0.062 &                         0.077 &                             0.059 \\
\bottomrule

\end{tabular}
}
}


\end{table*}

\clearpage

\section*{Acknowledgments}

 We would like to thank the technicians and the engineers who contributed to the design and construction of the CALICE AHCAL prototype detector. We also gratefully acknowledge the CERN management for its support and hospitality and its accelerator staff for the reliable and efficient operation of the test beam. The authors acknowledge the support from the BMBF via the High-D consortium. This work is supported by the Deutsche Forschungsgemeinschaft (DFG, German Research Foundation) under Germany's Excellence Strategy, EXC 2121, Quantum Universe (390833306).

\end{document}